\documentclass[review,12pt,authoryear]{elsarticle}

\usepackage{amssymb}
\usepackage{amsthm}
\usepackage{amsmath}
\usepackage{mathrsfs}
\usepackage{graphicx}
\usepackage{epstopdf}
\usepackage{float}
\usepackage{caption}
\usepackage{subcaption}
\usepackage{bm}
\usepackage{bbm}
\usepackage{mathrsfs}
\usepackage{cleveref}
\usepackage{soul}
\usepackage{tabu}
\usepackage{upgreek} 
\usepackage[dvipsnames]{xcolor}
\biboptions{sort&compress}

\makeatletter
\def\thickhline{%
  \noalign{\ifnum0=`}\fi\hrule \@height \thickarrayrulewidth \futurelet
   \reserved@a\@xthickhline}
\def\@xthickhline{\ifx\reserved@a\thickhline
               \vskip\doublerulesep
               \vskip-\thickarrayrulewidth
             \fi
      \ifnum0=`{\fi}}
\makeatother

\newlength{\thickarrayrulewidth}
\journal{International Journal of Plasticity}

\makeatletter
\def\@author#1{\g@addto@macro\elsauthors{\normalsize%
    \def\baselinestretch{1}%
    \upshape\authorsep#1\unskip\textsuperscript{%
      \ifx\@fnmark\@empty\else\unskip\sep\@fnmark\let\sep=,\fi
      \ifx\@corref\@empty\else\unskip\sep\@corref\let\sep=,\fi
      }%
    \def\authorsep{\unskip,\space}%
    \global\let\@fnmark\@empty
    \global\let\@corref\@empty  
    \global\let\sep\@empty}%
    \@eadauthor={#1}
}
\makeatother

\begin{document}

\begin{frontmatter}



\title{A mechanism-based multi-trap phase field model for hydrogen assisted fracture}


\author{Mehrdad Isfandbod\fnref{IC}}

\author{Emilio Mart\'{\i}nez-Pa\~neda\corref{cor1}\fnref{IC}}
\ead{e.martinez-paneda@imperial.ac.uk}

\address[IC]{Department of Civil and Environmental Engineering, Imperial College London, London SW7 2AZ, UK}

\cortext[cor1]{Corresponding author.}

\begin{abstract}
We present a new mechanistic, phase field-based formulation for predicting hydrogen embrittlement. The multi-physics model developed incorporates, for the first time, a Taylor-based dislocation model to resolve the mechanics of crack tip deformation. This enables capturing the role of dislocation hardening mechanisms in elevating the tensile stress, hydrogen concentration and dislocation trap density within tens of microns ahead of the crack tip. The constitutive strain gradient plasticity model employed is coupled to a phase field formulation, to simulate the fracture process, and to a multi-trap hydrogen transport model. The analysis of stationary and propagating cracks reveals that the modelling framework presented is capable of adequately capturing the sensitivity to the hydrogen concentration, the loading rate, the material strength and the plastic length scale. In addition, model predictions are compared to experimental data of notch tensile strength versus hydrogen content on a high-strength steel; a very good agreement is attained. We define and implement both atomistic-based and phenomenological hydrogen degradation laws and discuss similarities, differences and implications for the development of parameter-free hydrogen embrittlement models.
\end{abstract}


\end{frontmatter}

\section{Introduction}
\label{Introduction}

The ingress of hydrogen into a metal is known to cause a dramatic reduction in material strength, ductility, toughness and fatigue resistance \citep{Gangloff2003}. This phenomenon, referred to as hydrogen embrittlement, has attracted the attention of the material science and solid mechanics communities for decades due to its important technological implications and the scientific challenges inherent to its complex chemo-micromechanical nature \citep{Djukic2019}. Moreover, the problem has come very much to the fore in recent years as a consequence of the higher susceptibility of new, high-strength alloys, and because of the promise that hydrogen holds as a future energy carrier, requiring the development of suitable structures for hydrogen storage and transport \citep{Gangloff2012,Paxton2017}.\\

A notable effort has been devoted to the development of multi-physics models for predicting hydrogen assisted failures; despite the complexity of understanding and reproducing the underlying mechanisms, which span multiple scales and can vary significantly from one material to another (\citealp{Dadfarnia2010}; \citealp{Harris2018}; \citealp{Lynch2019}; \citealp{Shishvan2020}; \citealp{JMPS2020}). A wide variety of continuum-like models have been presented to predict and couple the multiple elements of this phenomenon: the deformation of the solid, the uptake and diffusion of hydrogen through the crystal lattice, and the resulting hydrogen assisted damage. The constitutive behaviour of the solid is typically characterised by using conventional von Mises plasticity theory, occasionally incorporating the effects of hydrogen-induced dilatation \citep{Lufrano1998a} and hydrogen-induced softening \citep{Diaz2016b}. Recently, there has been a growing interest in enriching the constitutive behaviour using strain gradient plasticity, so as to provide a more accurate description of crack tip fields by incorporating the influence of Geometrically Necessary Dislocations (GNDs) and plastic strain gradients (\citealp{AM2016}; \citealp{Kumar2020}). In regard to hydrogen transport, models based on Fickian diffusion have been developed to capture bulk transport \citep{VanLeeuwen1974}, and subsequently extended to capture stress-assisted diffusion and the role of microstructural traps in retaining hydrogen. These models take as primal kinematic variable either the lattice hydrogen concentration (\citealp{Sofronis1989}; \citealp{Barrera2016}; \citealp{AM2020}) or the chemical potential \citep{DiLeo2013,Elmukashfi2020}. In addition, generalised boundary conditions have been proposed to resolve the electrochemical-diffusion interface \citep{CS2020b}, in a first step to quantifying not only hydrogen diffusion but also ingress \citep{Kehler2008}. Finally, coupled deformation-diffusion models have been extended to explicitly predict hydrogen-assisted crack initiation and growth. The vast majority of these models are based on the concept of a fracture process zone, with hydrogen degrading the fracture energy of the solid. The most popular methodology has arguably been the use of cohesive zone models; see, e.g., (\citealp{Serebrinsky2004}; \citealp{Yu2016a}; \citealp{EFM2017}) and Refs. therein. However, phase field fracture models have recently been proposed to overcome the limitations intrinsic to cohesive zone formulations and other discrete methods. The phase field has emerged as a promising variational tool for fracture; enabling capturing - on the original finite element mesh and in arbitrary geometries and dimensions - complex cracking phenomena such as crack nucleation, branching, kinking or merging (\citealp{Bourdin2000}; \citealp{Duda2015}; \citealp{Miehe2016b}; \citealp{TAFM2020}; \citealp{CMAME2021}). This success has recently been extended to hydrogen embrittlement, with multi-physics phase field fracture formulations quickly gaining traction, and demonstrating their ability to reproduce experimental results and predict failures in service conditions (\citealp{CMAME2018}; \citealp{Duda2018}; \citealp{CS2020}; \citealp{Wu2020b}; \citealp{TAFM2020c}).\\

In this work, we present a new phase field-based formulation for hydrogen assisted cracking. The model incorporates, for the first time, the influence of hydrogen traps in a phase field framework. Multiple trap types are considered, capturing their influence on diffusion and fracture. Moreover, phase field fracture is coupled with a mechanism-based strain gradient plasticity model, also for the first time. This is of notable importance as it provides an enriched description of crack tip fields over the critical length scale for hydrogen damage. Consider for example the fracture experiments on Monel K500 reported in \citep{AM2016}; the crack growth rate in stage II ($\text{d}a/\text{d}t_{II}$), where cracking is intermittent and $\text{d}a/\text{d}t$ is constant (diffusion-controlled), is on the order of 0.04 $\upmu$m/s for three replicate experiments under an applied potential of $E_A=-1000$ m$V_{SCE}$. Given that the diffusion distance can be approximated as $d=2\sqrt{Dt}$, with Monel K500 having a diffusion coefficient $D=0.01$ $\upmu$m$^2$/s, this results in a critical distance of $x_{crit} \sim 1$ $\upmu$m. It is well known that conventional continuum models fail to capture the dislocation hardening mechanisms governing material deformation at the micro-scale. Of particular relevance to cracks in engineering components is the flow stress elevation associated with plastic strain gradients and large dislocation densities, as measured under similar conditions using indentation or a plethora of micro-scale experiments; from micro-torsion to constrained shear of thin films (\citealp{Fleck1994}; \citealp{Tvergaard2004a}; \citealp{Gurtin2005a}; \citealp{Mu2014}; \citealp{Voyiadjis2019}). The plastic zone adjacent to the crack tip is physically small and contains large gradients of plastic strain, leading to local strengthening and a stress elevation that can have important implications for hydrogen embrittlement, given the exponential dependence of the hydrogen concentration on hydrostatic stresses and the central role that crack tip stresses play in triggering interface decohesion (\citealp{Wei1997}; \citealp{Wei2005}; \citealp{Komaragiri2008}; \citealp{IJSS2015}; \citealp{IJP2016}). In addition, the dislocation trap density near the crack tip could be larger than that predicted by conventional continuum theories, due to the GNDs contribution. 
These features are captured here by means of a formulation based on \citet{Taylor1938} dislocation density model.\\

The remainder of this paper is organised as follows. In Section \ref{Sec:Theory} we present our theoretical framework. Then, the finite element implementation is described in Section \ref{Sec:NumModel}. Representative numerical results are shown in Section \ref{Sec:FEMresults}. First, a boundary layer model is used to gain insight into crack initiation and growth under small scale yielding conditions. Secondly, we compare the predictions of our model with experiments conducted on high-strength alloys exposed to hydrogenous environments. Concluding remarks end the paper in Section \ref{Sec:ConcludingRemarks}.\\

\noindent \textit{Notation.}
We use lightface italic letters for scalars, e.g. $\phi$, upright bold letters for vectors, e.g. $\mathbf{u}$, and bold italic letters, such as $\bm{\sigma}$, for second and higher order tensors. Inner products are denoted by a number of vertically stacked dots, corresponding to the number of indices over which summation takes place, such that $\bm{\sigma} : \bm{\varepsilon} = \sigma_{ij} \varepsilon_{ij}$, with indices referring to a Cartesian coordinate system. The gradient and the Laplacian are respectively denoted by $\nabla \mathbf{u}= u_{i,j}$ and $\Delta \phi= \phi_{,ii}$. Finally, divergence is denoted by $\nabla \cdot \bm{\sigma}=\sigma_{ij,j}$, the trace of a second order tensor is written as $\text{tr} \,\bm{\varepsilon}=\varepsilon_{ii}$, and the deviatoric part of a tensor is written as $\bm{\sigma}'=\sigma_{ij}-\delta_{ij} \sigma_{kk}/3$, with $\delta_{ij}$ denoting the Kronecker delta.

\section{Theory}
\label{Sec:Theory}

In this section, we formulate our theory, which couples deformation, fracture and hydrogen transport in strain gradient plasticity solids. The theory refers to the response of a body occupying an arbitrary domain $\Omega \subset {\rm I\!R}^n$ $(n \in[1,2,3])$, with external boundary $\partial \Omega\subset {\rm I\!R}^{n-1}$, on which the outwards unit normal is denoted as $\mathbf{n}$. In the following, we restrict our attention to isothermal conditions and isotropic solids.

\subsection{Kinematics}

We shall now discuss the independent fields that will be used to describe the kinematical structure of $\Omega$. In regard to the deformation of the solid, the motion of a material point $\mathbf{x}$ at a time $t$ is characterised by a displacement vector field $\mathbf{u} (\mathbf{x}, t)$. Such that, assuming a small strain formulation, the local deformation is determined by the infinitesimal strain tensor field $\bm{\varepsilon}$; given by
\begin{equation}
    \bm{\varepsilon} = \frac{1}{2}\left(\nabla\mathbf{u}^T+\nabla\mathbf{u}\right).
\end{equation}

Our constitutive theory considers both elastic and plastic strains, adopting the standard partition:
\begin{equation}\label{eq:E=EeEp}
    \bm{\varepsilon} = \bm{\varepsilon}^e + \bm{\varepsilon}^p \, .
\end{equation}

The nucleation and growth of cracks are described by means of a phase field variable, or order parameter, $\phi (\mathbf{x}, t) \in [0, 1]$. Following standard damage mechanics arguments, the phase field equals $\phi=0$ for intact material and $\phi=1$ for fractured material points. Using a phase field auxiliary variable to \emph{implicitly} track interfaces has opened new horizons in the modelling of fracture \citep{Wu2020}, microstructural evolution \citep{Provatas2011}, and metal corrosion \citep{JMPS2021}. Here, in the context of fracture, the phase field variable must grow monotonically,
\begin{equation}\label{eq:PhiGrowthMonotonically}
    \dot{\phi} (\mathbf{x}, t) \geq 0 ,
\end{equation}

\noindent so as to ensure that the microstructural changes associated with damage are irreversible.\\

In addition, we consider the changes in composition of material points in $\Omega$. The presence of guest hydrogen atoms in a host metallic lattice is characterised by the hydrogen concentration $C (\mathbf{x}, t)$. Here, $C (\mathbf{x}, t)$ denotes the total number of moles of hydrogen atoms per unit reference volume. Even though mass concentration is the sought variable, the thermodynamic driving force for diffusion is the chemical potential gradient $\nabla \mu$. As in \citet{Duda2018}, we define a scalar field $\eta (\mathbf{x}, t)$ to determine the kinematics of composition changes, such that
\begin{equation}
    \dot{\eta} = \mu \,\,\,\,\,\,\, \text{and} \,\,\,\,\,\,\, \eta (\mathbf{x}, t) = \int_0^t \mu (\mathbf{x}, t) \, \text{d}t \, .
\end{equation}

\subsection{Principle of virtual work. Balance of forces}

We shall now derive the balance equations for the coupled system using the principle of virtual work. Consider the three-field boundary value problem resulting from the primal kinematic variables described before: $\mathbf{u}$, $\phi$, and $\eta$. The Cauchy stress $\bm{\sigma}$ is introduced, which is work conjugate to the strain tensor $\bm{\varepsilon}$. Correspondingly, for an outwards unit normal $\mathbf{n}$ on the boundary $\partial\Omega$ of the solid, a vector traction $\mathbf{T}$ is defined, which is work conjugate to the displacement field $\mathbf{u}$. The damage response involves a scalar stress-like quantity $\omega$, which is work conjugate to the phase field $\phi$, and a phase field micro-stress vector $\bm{\upxi}$ that is work conjugate to the gradient of the phase field $\nabla\phi$. The phase field is assumed to be driven by the displacement problem alone; i.e., no external traction is associated with $\phi$. In regard to the mass transport, the surface flux is denoted by $\mathbf{J}$, and accordingly, a concentration flux entering the body across $\partial \Omega$ can be defined as $q=\mathbf{J} \cdot \mathbf{n}$. Then, making use of three virtual fields ($\delta \mathbf{u}$, $\delta \phi$, $\delta \eta$), the principle of virtual work for the coupled system reads,
\begin{equation}\label{eq:PVW}
  \int_\Omega  \left( \bm{\sigma} : \delta \bm{\varepsilon}  + \omega \, \delta \phi + \bm{\upxi} \cdot \nabla \delta \phi - \dot{C} \, \delta \eta + \mathbf{J} \cdot \nabla \delta \eta \right) \, \text{d} V = \int_{\partial \Omega} \left( q \, \delta \eta +\mathbf{T} \cdot \delta \mathbf{u} \right) \, \text{d} S    
\end{equation}

The principle of virtual work must hold for an arbitrary domain $\Omega$ and for any kinematically admissible variations of the virtual quantities. Thus, by making use of Gauss's divergence theorem, the local force balances are obtained as: 
\begin{equation}
    \begin{split}
        &\nabla\cdot\bm{\sigma}=0  \\
        &\nabla\cdot\bm{\upxi}-\omega =0 \\
        &\dot{C} + \nabla\cdot \mathbf{J}=0
    \end{split}\hspace{2cm} \text{in } \Omega,\label{eq:balance}
\end{equation}
\noindent with natural boundary conditions: 
\begin{equation}
    \begin{split}
        \bm{\sigma}\cdot\mathbf{n}=\mathbf{T} \\
         \bm{\upxi} \cdot \mathbf{n}=0 \\
        q= \mathbf{J}\cdot\mathbf{n}
    \end{split} \hspace{2cm} \text{on } \partial\Omega.\label{eq:balance_BC}
\end{equation}

\subsection{Energy imbalance}

We shall now impose the first and second laws of thermodynamics through an energy imbalance. The first two laws of thermodynamics for a continuum body within a dynamical process of specific internal energy $\mathscr{E}$ and specific entropy $\mit \Lambda$ read \citep{Gurtin2010},
\begin{equation}\label{eq:thermodynamics laws}
    \begin{aligned}
    & \frac{\text{d}}{\text{d}t} \int_\Omega \mathscr{E} \, \text{d} V = \dot W_e \left( \Omega \right) - \int_{\partial \Omega} \mathbf{Q} \cdot \mathbf{n} \, \text{d} S + \int_\Omega Q \, \text{d} V \\
    & \frac{\text{d}}{\text{d}t} \int_\Omega {\mit \Lambda} \, \text{d} V \geqslant - \int_{\partial \mathrm{\Omega}} \frac{ \mathbf{Q} }{T} \cdot \mathbf{n} \, \text{d} S +  \int_\mathrm{\Omega} \frac{Q}{T} \, \text{d} V \, .
    \end{aligned} 
\end{equation}

\noindent Here, $\dot W_e$ is the power of external work, $\mathbf{Q}$ is the heat ﬂux, and $Q$ is the heat absorption. The thermodynamic laws in the presence of species transport require that the temporal increase in free energy of any part $\Omega$ is less than or equal to the power expended on $\Omega$ plus the flux of energy carried into $\Omega$ through its boundary $\partial \Omega$ by the diffusing species (see \citealp{Anand2019}). Denoting $\psi$ as the free energy per unit reference volume, this constraint takes the form of the following free energy imbalance,
\begin{equation}\label{eq:Clausius}
    \frac{\text{d}}{\text{d}t} \int_\Omega \psi \, \text{d}V \leq \int_{\partial\Omega}\dot{W}_{e} \, \text{d}S + \int_{\partial \Omega} \mu \, \mathbf{J} \cdot \mathbf{n} \, \text{d} S \, .
\end{equation}

Consider now the local balance equations and recall that the external work is given by the right-hand side of (\ref{eq:PVW}). Applying the divergence theorem to the last term in (\ref{eq:Clausius}), considering the strain partitioning (\ref{eq:E=EeEp}) and replacing virtual fields ($\delta a$) by realisable velocity fields ($\dot{a}$), one reaches:
\begin{equation}\label{eq:FreeEnergyImbalance1}
    \int_\Omega \left[ \dot{\psi} - \left( \bm{\sigma} : \dot{\bm{\varepsilon}}^e +\bm{q} : \dot{\bm{\varepsilon}}^p + \omega \dot{\phi} + \bm{\upxi} \cdot \nabla \dot{\phi} \right) + \mu \dot{C} - \mathbf{J} \cdot \nabla \mu \right] \text{d}V \leq 0 \, ,
\end{equation}

\noindent where a plastic micro-stress tensor $\bm{q}$ has been defined (work-conjugate to $\bm{\varepsilon}^p$). Since (\ref{eq:FreeEnergyImbalance1}) must hold for any volume $\Omega$, it follows that it must also hold in a local fashion, such that the local free-energy imbalance reads,
\begin{equation}
    \dot{\psi} - \bm{\sigma} : \dot{\bm{\varepsilon}} - \omega \dot{\phi} - \bm{\upxi} \cdot \nabla \dot{\phi} + \mu \dot{C} - \mathbf{J} \cdot \nabla \mu \leq 0 \, .
\end{equation}

Accordingly, appropriate constitutive relations must be considered in order to fulfil the following imbalance:
\begin{align}\label{eq:FreeEnergyImbalance2}
  \left( \bm{\sigma}- \dfrac{\partial\psi}{\partial\bm{\varepsilon}^e} \right)  : \dot{\bm{\varepsilon}^e}  & +  \left( \bm{q}- \dfrac{\partial\psi}{\partial\bm{\varepsilon}^p} \right)  : \dot{\bm{\varepsilon}^p} +\left(\omega-\dfrac{\partial\psi}{\partial\phi}\right)\dot{\phi}  + \left( \bm{\upxi}-\dfrac{\partial\psi}{\partial\nabla\phi} \right) \cdot\nabla\dot{\phi} \nonumber \\
 & - \left[ \left( \mu-\dfrac{\partial\psi}{\partial C} \right) \dot{C} - \mathbf{J} \cdot \nabla \mu \right] \geq 0,   
\end{align}

\subsection{Constitutive theory}

We shall now proceed to develop a constitutive theory for the coupled deformation-diffusion-fracture problem that is consistent with the free energy imbalance, Eq. (\ref{eq:FreeEnergyImbalance2}). 

\subsubsection{Mechanism-based elasto-plasticity}

We start by outlining the constitutive choices that characterise the microstructural changes leading to plasticity and fracture. First, the influence of damage into the mechanical deformation of the body is captured by defining a deformation free energy that decreases with the phase field variable $\phi$. Specifically, the following quadratic degradation function is chosen:
\begin{equation}\label{eq:degradationfunction}
    g \left( \phi \right) = \left( 1 - \phi \right)^2 \, .
\end{equation}

\noindent Thus, the relation between the Cauchy stress tensor $\bm{\sigma}$ and the undamaged stress tensor $\bm{\sigma}_0$ follows immediately. Omitting the negligible role of lattice dilation \citep{Hirth1980} and consistent with (\ref{eq:FreeEnergyImbalance2}), the stress tensor is given by
\begin{equation}\label{eq:Cauchy}
\bm{\sigma} = \frac{\partial \psi}{\partial \bm{\varepsilon}^e} = \left( 1 - \phi \right)^2 \bm{\sigma}_0 = \left( 1 - \phi \right)^2 \bm{C}^e : \bm{\varepsilon}^e = \left( 1 - \phi \right)^2 \bm{C} : \bm{\varepsilon} \, ,
\end{equation}

\noindent where $\bm{C}^e$ denotes the elastic stiffness tensor and $\bm{C}$ is the consistent elastic-plastic material Jacobian. As discussed in the introduction, the latter is defined in a suitable manner to capture the important role that dislocation hardening mechanisms associated with GNDs and plastic strain gradients play on crack tip mechanics. For this, a mechanism-based formulation is developed, grounded on \citet{Taylor1938} dislocation model \citep{Huang2004a,Liu2005}.\\

Considering \citet{Taylor1938} model as underlying principle, the shear flow stress $\tau$ is formulated in terms of the dislocation density $\rho$ as
\begin{equation}\label{eq:1MSG}
    \tau = \alpha G b \sqrt{\rho} \, ,
\end{equation}

\noindent where $\alpha$ is an empirical coefficient taking values between 0.3 and 0.5, $G$ denotes the shear modulus, and $b$ corresponds to the Burger's vector. The dislocation density can be additively decomposed into the sum of the density of statistically stored dislocations (SSDs) $\rho_S$, which trap each other in a random way, and the density of geometrically necessary dislocations (GNDs) $\rho_G$, which are required for the compatible deformation of the crystal. Hence, 
\begin{equation} \label{Eq:2MSG}
\rho = \rho_S + \rho_G  \, .
\end{equation}
The GND density $\rho_G$ is related to the effective plastic strain gradient $\eta^{p}$ by: 
\begin{equation} \label{Eq:3MSG}
\rho_G = \overline{r}\frac{\eta^{p}}{b} \, ,
\end{equation}
\noindent where $\overline{r}$ is the Nye-factor, which is assumed to be approximately 1.90 \citep{Arsenlis1999,Shi2004}. Following \citet{Fleck1997}, three quadratic invariants of the plastic strain gradient tensor are used to represent the effective plastic strain gradient $\eta^{p}$ as
\begin{equation}
\eta^{p}=\left( c_1 \eta^{p}_{iik} \eta^{p}_{jjk} + c_2 \eta^{p}_{ijk} \eta^{p}_{ijk} + c_3 \eta^{p}_{ijk} \eta^{p}_{kji} \right)^{1/2} \, .
\end{equation}

The coefficients have been determined to be equal to $c_1=0$, $c_2=1/4$ and $c_3=0$ from three dislocation models for bending, torsion and void growth \citep{Gao1999}, leading to
\begin{equation}
\eta^{p}=\frac{1}{2}\left( \bm{\eta}^{p} \bm{\eta}^{p} \right)^{1/2} \, ,
\end{equation}

\noindent where the components of the strain gradient tensor are obtained by $\eta^{p}_{ijk}= \varepsilon^{p}_{ik,j}+\varepsilon^{p}_{jk,i}-\varepsilon^{p}_{ij,k}$. The tensile flow stress $\sigma_{f}$ is related to the shear flow stress $\tau$ through the Taylor factor $M$, such that
\begin{equation} \label{eq:4MSG}
\sigma_{f} =M\tau = M \alpha G b \sqrt{\rho} \, .
\end{equation}

This linear dependence of the square of plastic flow stress on strain gradients, resulting from \citet{Taylor1938} dislocation model, is intrinsic to the mechanism-based strain gradient (MSG) plasticity theory and is grounded on the nano-indentation experiments by \citet{Nix1998}. 
Rearranging Eqs. (\ref{eq:1MSG}-\ref{Eq:3MSG}) and substituting into (\ref{eq:4MSG}), the flow stress can be re-formulated as
\begin{equation} \label{Eq5MSG}
\sigma_{f} =M\alpha G b \sqrt{\rho_{S}+\overline{r}\frac{\eta^{p}}{b}} \, .
\end{equation}

The SSD density $\rho_{S}$ can be determined from (\ref{Eq5MSG}) knowing the relation in uniaxial tension between the flow stress and the material stress-strain curve as follows
\begin{equation} \label{Eq:6MSG}
\rho_{S} = \left( \frac{\sigma_{ref}f(\varepsilon^{p})}{M\alpha G b} \right)^2
\end{equation}

\noindent Here, $\sigma_{ref}$ is a reference stress and $f$ is a non-dimensional function of the equivalent plastic strain $\varepsilon^{p}$, as given from the uniaxial stress-strain curve. Substituting back into (\ref{Eq5MSG}), $\sigma_{f}$ yields:
\begin{equation} \label{Eq:Sflow}
\sigma_{f} =\sigma_{ref} \sqrt{f^2(\varepsilon^{p})+L_p \eta^{p}}
\end{equation}

\noindent where $L_p$ is the intrinsic plastic material length. It can be readily seen that conventional von Mises plasticity is recovered if $L_p=0$ or if the gradient-related term $L_p \eta^{p}$ becomes negligible relative to the characteristic length of plastic deformation. Also, we emphasise that the theory is intended to model a collective behaviour of dislocations, implying that it is only applicable at a much larger scale than the average dislocation spacing; i.e. distances of 100 nm or larger. This is also the scale over which the differences between higher order and lower order versions of mechanism-based strain gradient plasticity are relevant \citep{Shi2001}. I.e., identical results are expected over the regime where continuum models are applicable - a distance of 100 nm or larger ahead of the crack tip. Thus, we choose to adopt a lower order formulation by using a viscoplastic approach, following \citet{Huang2004a}. The purpose is to overcome the need for higher order stresses by constructing a self-contained model through the viscoplastic relation between $\dot{\varepsilon}^p$ and the effective von Mises stress $\sigma_e$. To mimic a rate-independent response, we use the viscoplastic-limit approach by \citet{Kok2002}, which entails replacing the reference strain rate $\dot{\varepsilon}_0$ with the effective strain rate $\dot{\varepsilon}$. Accordingly, the effective plastic strain rate is defined as,
\begin{equation}
    \dot{\varepsilon}^p = \dot{\varepsilon} \left[ \frac{\sigma_e}{\sigma_{ref} \sqrt{f^2 \left(\varepsilon^p \right) + L_p \eta^{p}}} \right]^m \, ,
\end{equation}

\noindent where $m$ is the strain rate sensitivity exponent; values larger than 5 provide a response similar to that of rate-independent solids, with no differences being observed for $m \geq 20$ \citep{Huang2004a}. A value of $m=20$ is adopted in this work. The consistent material Jacobian is then given by \citep{Qu2004,TAFM2017},
\begin{equation}\label{eq:MaterialJacobian}
   \bm{\sigma}_0= K \text{tr} \left( \dot{\bm{\varepsilon}} \right) \bm{\delta} + 2 \mu \left\{ \dot{\bm{\varepsilon}}' - \frac{3 \dot{\varepsilon}}{2 \sigma_e} \left[ \frac{\sigma_e}{\sigma_{ref} \sqrt{f^2 \left( \varepsilon^p \right) + L_p \eta^p}} \right]^m \right\} .
\end{equation}

\noindent Here, $K$ is the bulk modulus and $\bm{\delta}$ is the Kronecker delta. Finally, the plastic hardening behaviour is given by the following isotropic hardening power law:
\begin{equation}\label{eq:PoweLaw}
    \sigma = \sigma_Y \left( 1 + \frac{E \varepsilon^p}{\sigma_Y} \right)^N
\end{equation}

\noindent where $N$ is the strain hardening exponent ($0\leq N \leq 1$). Accordingly, $\sigma_{ref}=\sigma_Y ( E/ \sigma_Y )^N$ and $f (\varepsilon^p) = (\varepsilon^p + \sigma_Y / E)^N $. The mechanical constitutive response of the solid is then completely characterised by the damage degradation function (\ref{eq:degradationfunction}), the Cauchy stress definition (\ref{eq:Cauchy}) and the consistent material Jacobian (\ref{eq:MaterialJacobian}).

\subsubsection{Hydrogen-sensitive phase field damage}
\label{Sec:PhaseFieldTheory}

We proceed to describe the fracturing process by following and extending the phase field model for hydrogen embrittlement developed by \citet{CMAME2018}. The idea is to approximate the fracture energy density $\psi^f$ over a discontinuous surface $\Gamma$ by using a smooth and continuous auxiliary (phase) field that smears the crack. The model is non-local and involves as a consequence a phase field length scale $\ell$, which governs the smearing of the crack. It has been shown using Gamma-convergence that the regularised fracture energy converges to the original form for a vanishing value of $\ell$ \citep{Bellettini1994,Chambolle2004}. Accordingly, for a material of toughness $G_c$,
\begin{equation}
    \psi^f = \int_\Gamma G_c \, \text{d}S \approx \int_\Omega G_c \gamma \left( \phi, \, \nabla \phi \right) \, \text{d}V \, .
\end{equation}

\noindent Here, $\gamma$ is the crack density functional, which is here chosen to be:
\begin{equation}
   \gamma \left( \phi, \, \nabla \phi \right) = \frac{\phi^2}{2 \ell} + \frac{\ell}{2} |\nabla \phi|^2 \, . 
\end{equation}

This phase field approximation circumvents the need to track discrete crack surfaces, enabling the modelling of complex fracture phenomena (see, e.g., \citealp{McAuliffe2015}; \citealp{CPB2019}; \citealp{Quintanas-Corominas2020a}).\\

Hydrogen comes into the picture by degrading the toughness of the solid, as consistently observed in laboratory experiments \citep{Gangloff2003}. The specific degradation law can be chosen in a phenomenological manner or by establishing a connection with the underlying physical micromechanisms. Both options will be explored here. In its most general form, the evolution of the critical fracture energy can be given by,
\begin{equation}\label{eq:Gc}
    G_c (C) = f \left( C \right) G_c (0) \, ,
\end{equation}

\noindent where $f \left( C \right)$ is a hydrogen degradation function. In this work, the focus will be on alloys that exhibit intergranular fracture in the presence of hydrogen. Thus, we enrich (\ref{eq:Gc}) by defining $f (C)$ to be a function of the hydrogen trapped at the grain boundaries, as predicted by the multi-trap hydrogen transport model described below.\\

It remains to define the constitutive relations for the micro-stress variables work conjugate to the phase field and the phase field gradient. Denoting by $\psi^b$ the bulk strain energy density of the solid, the scalar micro-stress $\omega$ is given by,
\begin{equation}\label{eq:consOmega}
    \omega = \dfrac{\partial\psi}{\partial\phi} = -2(1-\phi)\psi^b+G_c (C) \dfrac{\phi}{\ell}.
\end{equation}

\noindent Likewise, the phase field micro-stress vector $\bm{\upxi}$ reads:
\begin{equation}\label{eq:consXi}
    \bm{\upxi} = \dfrac{\partial\psi}{\partial\nabla\phi} = G_c (C) \, \ell \, \nabla\phi.
\end{equation}

Accordingly, the phase field local balance (\ref{eq:balance}c) can be re-formulated by considering the constitutive choices (\ref{eq:consOmega}) and (\ref{eq:consXi}). Assuming a zero concentration gradient along the small region where $\nabla \phi \neq 0$, the phase field evolution equation reads:
\begin{equation}\label{eq:PhaseFieldStrongForm}
  G_c ( C) \left( \frac{\phi}{\ell} - \ell \nabla^2 \phi \right) - 2 (1 - \phi) \psi^b = 0  \, .
\end{equation}

Fracture is assumed to be driven by the bulk strain energy density of the solid, which is here defined as the summation of the elastic and plastic strain energy densities:
\begin{equation}
    \psi^b = \psi^e + \psi^p = \frac{1}{2} \bm{\varepsilon}^e : \bm{C}^e : \bm{\varepsilon}^e + \int_0^t \left( \bm{\sigma} : \dot{\bm{\varepsilon}}^p \right) \, \text{d}t \, .
\end{equation}

\noindent where $\bm{C}^e$ is the linear elastic stiffness matrix. Thus, both elastic and plastic strain energy densities contribute on an equal footing to the fracture process, as in e.g. (\citealp{Miehe2016b}; \citealp{CS2020}). Other approaches have also been proposed, including a driving force for fracture based purely on the elastic stored energy \citep{Duda2015,Duda2018} or the consideration of elastic and plastic energies with a different weighting (see, e.g., \citealp{Borden2016,You2021}). A physically-sound choice is not straightforward as it is unclear to what extent a Griffith-type energy balance applies to ductile fracture \citep{Orowan1948,Hutchinson1983}.

\subsubsection{A multi-trap model for hydrogen transport}

Our theory deals with the dilute transport of hydrogen in metals. Hydrogen atoms occupy normal  interstitial lattice sites (NILS) and can additionally reside at trapping sites such as interfaces or dislocations. We adopt the common assumptions of the literature (see, e.g., \citealp{Sofronis1989,DiLeo2013}) and base our modelling on the equilibrium theory presented by Oriani \citep{Oriani1974}. The subscript $L$ refers to lattice sites and the subscript $T$ to trap sites. Superscripts are used to denote the different trap sites. It is assumed that traps are isolated (i.e., do not form an extended network). Hence, hydrogen transport between trap sites is by lattice diffusion.\\

The hydrogen concentration in lattice sites is given by,
\begin{equation}\label{Eq:CL}
C_L=N_L \beta \, \theta_L
\end{equation}

\noindent where $N_L$ is the density of the host metal lattice measured in solvent atoms per unit volume, $\beta$ is the number of interstitial sites per atom, and $\theta_L$ is the lattice occupancy fraction ($0 \leq \theta_L \leq 1$). The number of interstitial sites per solvent atom $\beta$ is typically taken to be equal to 6 for bcc metals, as indirect evidence suggests that tetrahedral site occupancy is favoured relative to octahedral site occupancy at room temperature \citep{Hirth1980,Kiuchi1983}. For fcc lattices, $\beta=1$ is usually assumed, resulting from the more favourable octahedral site occupancy. The density of solvent atoms is a function of the molar volume of the host lattice $V_M$ and Avogadro's number $N_A$ as,
\begin{equation}\label{Eq:NLVA1}
N_L = \frac{N_A}{V_M}=\frac{N_A \rho_M}{M_M} \, ,
\end{equation}

\noindent where $\rho_M$ is the density and $M_M$ is the molar mass. In the case of iron at 293 K, the density equals $\rho_M=7.87 \times 10^3$ kg/m$^3$ and the atomic weight $M_M=55.8 \times 10^{-3}$ kg/mol; this implies $N_L=8.46 \times 10^{28}$ sites/m$^3$.\\ 

Hydrogen can also be retained at so-call hydrogen \emph{traps}. These are typically microstructural defects such as dislocations, grain boundaries, voids, carbides and interfaces. These traps can be reversible or irreversible, and can also be classified as saturable or unsaturable. Reversible traps are those that can immobilize and release hydrogen while irreversible traps are those that absorb hydrogen and prevent it from escaping. However, one should note that the term irreversible is not fundamentally correct but rather pragmatic, as leakage can always take place for a sufficiently long timescale or a sufficiently high temperature \citep{Turnbull2015}. Multiple trap types are considered here - the hydrogen concentration in the $i$th type of trapping site can be defined as:
\begin{equation}\label{Eq:CT}
C_T^{(i)} = \theta_T^{(i)} \alpha^{(i)} N_T^{(i)}
\end{equation}

\noindent where $N_T$ is the trap density, $\alpha$ is the number of atom sites per trap and $\theta_T$ is the fraction of occupied trapping sites. Hence, $\alpha N_T$ is the number of trapping sites per unit volume. We choose to adopt $\alpha=1$ and use apparent binding energies. The total hydrogen concentration is, therefore, the sum of the lattice hydrogen concentration and the concentration at each of the trap types considered:
\begin{equation}\label{Eq:SumC}
C =  C_L + \sum_i^n C_T^{(i)} \, .
\end{equation}

\noindent where $n$ is the total number of trap types.\\

The relation between the lattice and trapped hydrogen concentrations is assumed here following Oriani's equilibrium theory \citep{Oriani1974}. Thus, there is a Fermi-Dirac relation between the occupancy of the $i$th type of trapping sites and the fraction of occupied lattice sites
\begin{equation}\label{eq:Oriani}
    \frac{\theta_T^{(i)}}{1 - \theta_T^{(i)}} = \frac{\theta_L}{1- \theta_L} K^{(i)} \, ,
\end{equation}

\noindent with $K^{(i)}$ being the equilibrium constant for the $i$th type of trap, given by
\begin{equation}\label{eq:KT}
    K^{(i)}=\exp \left( \frac{-W_B^{(i)}}{RT} \right) \, .
\end{equation}

\noindent Here, $R$ is the gas constant, $T$ is the absolute temperature, and $W_B$ is the binding energy - an inherently negative quantity that quantifies the energy required for a hydrogen atom to escape a trap site and move into a lattice site. In many alloys, especially in bcc lattices, conditions of low occupancy $\theta_L << 1$ are usually assumed, such that
\begin{equation}\label{Eq:OrianiOccupancy}
\frac{\theta_L}{1-\theta_L} \approx \theta_L
\end{equation}

Considering (\ref{Eq:CL}), (\ref{Eq:CT}) and (\ref{eq:Oriani}), then the relation between the lattice and trapped concentration is established as,
\begin{equation}\label{Eq:CT2}
C_T^{(i)}=\frac{K^{(i)} \alpha^{(i)} N_T^{(i)}}{\beta N_L + \left(K^{(i)} -1 \right)C_L} C_L
\end{equation}

Mass diffusion is driven by gradients of chemical potential $\nabla \mu$. The mass flux is related to $\nabla \mu$ through a linear Onsager relation, grounded on Einstein's equation of diffusion. Accordingly, for a material with a diffusion coefficient $D$, the flux reads
\begin{equation}\label{Eq:JL1}
\mathbf{J}=-\frac{D}{RT} C_L \nabla \mu \, .
\end{equation}

Thus, hydrogen atoms migrate from regions of high chemical potential to regions of low chemical potential. Note that, for simplicity, the $L$ subscript is omitted from $\mathbf{J}$ and $\mu$ but both are related to the transport of hydrogen between interstitial lattice sites. The chemical potential is defined as,
\begin{equation}\label{Eq:MuL}
\mu = \mu^0 + RT \, \textnormal{ln} \frac{\theta_L}{1-\theta_L} - \bar{V}_H \sigma_H \, ,
\end{equation}

\noindent where $\mu^0$ denotes the chemical potential in the standard state, $\sigma_H$ is the hydrostatic stress, and $\bar{V}_H$ is the partial molar volume of hydrogen in solid solution ($\bar{V}_H=2000$ mm$^3$/mol for iron-based materials). As evident from (\ref{Eq:MuL}), the role of hydrostatic tensile stresses (volumetric strains) is to lower the chemical potential, increasing the hydrogen solubility in the lattice as a result of lattice dilatation.\\

Substituting (\ref{Eq:MuL}) into (\ref{Eq:JL1}) and adopting the common approximation of a constant interstitial sites concentration ($\nabla N_L=0$) gives:
\begin{equation}\label{Eq:JL2}
\mathbf{J}=-D \nabla C_L + \frac{D}{RT} C_L \bar{V}_H \nabla \sigma_H
\end{equation}

Fluxes, due to the chemical potential gradient, and hydrogen concentrations are related through the requirement of mass conservation:
\begin{equation}\label{Eq:MassBal}
\frac{\textnormal{d}}{\textnormal{d}t} \int_V C \,\, \textnormal{d}V + \int_S \mathbf{J} \cdot \mathbf{n} \,\, \textnormal{d} S=0
\end{equation}

Exploiting Oriani's equilibrium, an effective diffusion coefficient $D_e$ can be defined for a multi-trap system, such that the $D/D_e$ ratio reads,
\begin{equation}\label{eq:De}
    \frac{D}{D_e} = 1 + \sum_i \frac{\partial C_T^{(i)}}{\partial C_L} = 1 + \sum_i \left( \frac{K^{(i)} \alpha^{(i)} N_T^{(i)}/ \left( \beta N_L \right)}{\left[ 1 + \left( K^{(i)} -1 \right) C_L / \left( \beta N_L \right) \right]^2} \right)
\end{equation}

Now, consider Eqs. (\ref{Eq:SumC}), and (\ref{Eq:JL2})-(\ref{eq:De}). Making use of the divergence theorem and noting that (\ref{Eq:MassBal}) must hold for any arbitrary volume, the local mass balance can be derived as,
\begin{equation}\label{eq:StrongCL}
 \frac{D}{D_e}\frac{\partial C_L}{\partial t}=D \nabla^2 C_L-\nabla \left( \frac{D C_L}{RT} \bar{V}_H \nabla \sigma_H  \right ) \, .   
\end{equation}

It remains to define a constitutive choice for the trap density $N_T^{(i)}$. The trap density is often a material property that remains constant throughout the analysis, as it is the case for traps such as carbides or grain boundaries. However, dislocation traps evolve with mechanical loading and thus a constitutive law must be defined for $N_T^{(d)}$. We build upon the Taylor-based formulation presented above to establish an evolution law for $N_T^{(d)}$ in terms of the total dislocation density $\rho$, including both contributions from SSDs and GNDs. By assuming one trap site per atomic plane threaded by a dislocation, the following relation between the dislocation density $\rho$ and the trap site density $N_T^{(d)}$ can be identified:
\begin{equation}\label{Eq:NTrho}
N_T^{(d)}= \frac{1}{b} \rho
\end{equation}

\noindent where the pre-factor is the inverse of the Burgers vector, as slip occurs along the plane of the shortest Burgers vector. The total dislocation density can be computed from Eqs. (\ref{Eq:2MSG}), (\ref{Eq:3MSG}) and (\ref{Eq:6MSG}). Denoting $a$ the lattice parameter, the Burgers vector is given by $b=\sqrt{2}/a=0.2555$ nm for fcc metals, as slip occurs along the closed packed plane ${111}$ and slip direction $\langle \bar{1}10 \rangle$. For bcc metals $b=2/(\sqrt{3}a)=0.2725$ nm, as slip is assumed to occur along the $\{110\}$ plane and $\langle 111 \rangle$ direction. The reader is referred to, e.g., \citep{Davey1925} for a list of lattice constants for various metals.

\section{Numerical implementation}
\label{Sec:NumModel}

We shall now describe the numerical implementation of our coupled theory, in the context of the finite element method. First, the elastic strain energy density is decomposed to prevent the evolution of damage under compressive loading (Section \ref{subsec:Split}). Secondly, in Section \ref{subsec:History}, a history field is defined to prevent phase field damage reversibility. Finally, in Section \ref{subsec:DisFEM}, we address the discretisation of the weak formulation of our theory and formulate the residuals and the stiffness matrices. The implementation is conducted within an Abaqus user-element (UEL) subroutine, with the pre-processing of the input files carried out using Abaqus2Matlab \citep{AES2017}.

\subsection{Addressing damage in compression and crack interpenetration}
\label{subsec:Split}

Several formulations have been proposed to effectively decompose the elastic fracture driving force into tension and compression components, so as to prevent the nucleation of cracks under compressive stresses. Here, we follow the spherical/deviatoric split proposed by \citet{Amor2009}. Thus, in a solid with Lame's first parameter $\lambda$, the elastic strain energy density can be decomposed as $\psi^e=\psi_+^e+\psi_{-}^e$, with
\begin{equation}
    \begin{split}
        \psi_+^e &= \frac{1}{2}\left(\lambda +\frac{2}{3}G \right) \left\langle \text{tr }\bm{\varepsilon}^e\right\rangle_+^2 + G \, |{\bm{\varepsilon}^e}^{'}|^2\\
        \psi_{-}^e&=\frac{1}{2}\left(\lambda +\frac{2}{3}G \right) \left\langle \text{tr }\bm{\varepsilon}^e\right\rangle_-^2,
    \end{split}
\end{equation}

\noindent and only $\psi_+^e$ contributing to damage. Here, $\left\langle \right\rangle$ denotes the Macaulay brackets. The strain energy decomposition is implemented by means of a hybrid approach, following \citet{Ambati2015}. This implies that the split of the elastic strain energy density is incorporated into the phase field force balance but not considered in the balance of linear momentum. In addition, crack interpenetration is prevented by adding the following constraint \citep{Ambati2015}
\begin{equation}
    \phi=0 \hspace{1cm} \text{if } \psi_+^e < \psi_{-}^e.
\end{equation}

\subsection{Damage irreversibility}
\label{subsec:History}

Additional numerical restrictions are needed to ensure damage irreversibility, Eq. (\ref{eq:PhiGrowthMonotonically}). These constraints are restricted to the evolution of the elastic strain energy density, as it is assumed that the effective plastic work is monotonically increasing. We follow \citet{Miehe2010a} and introduce a history variable field $\mathcal{H}$. To ensure irreversible growth of the phase field variable, the history field must satisfy the Kuhn-Tucker conditions:
\begin{equation}
    \psi^e_+ - \mathcal{H} \leq 0 \, , \,\,\,\,\,\,\,\, \dot{\mathcal{H}} \leq 0 \, , \,\,\,\,\,\,\,\, \dot{\mathcal{H}} \left( \psi_+^e - \mathcal{H} \right)=0
\end{equation}

Thus, for a total time $t_t$, the history variable at time $t$ corresponds to the maximum value of $\psi_+^e$, i.e.:
\begin{equation}
    \mathcal{H} = \max_{t \in[0,t_t]}\psi^e_+( t). 
\end{equation}

\subsection{Finite element discretisation}
\label{subsec:DisFEM}

We proceed now to describe the finite element discretisation and the formulation of the residuals and stiffness matrices. Our numerical implementation uses as nodal unknowns the following fields: displacement $\mathbf{u}$, phase field $\phi$, and lattice hydrogen concentration $C_L$. We derive the weak form of the balance equation for each of these fields considering the constitutive choices outlined in Section \ref{Sec:Theory}. Neither body forces nor external tractions are considered, for simplicity. Thus, recalling (\ref{eq:PVW}) and (\ref{eq:degradationfunction}), the weak form for the mechanical problem reads,
\begin{equation}
  \int_\Omega \left\{ \left[ \left( 1 - \phi \right)^2 + k \right] \bm{\sigma}_0 : \delta \bm{\varepsilon} \right\} \text{d} V = 0   \, .
\end{equation}

\noindent where $k$ is a small positive-valued constant that is introduced to prevent ill-conditioning when $\phi=1$; a value of $k=1 \times 10^{-7}$ is chosen throughout this study. Regarding the phase field, the weak form is derived upon considering the history field $\mathcal{H}$ described above and combining (\ref{eq:PVW}) and (\ref{eq:PhaseFieldStrongForm}), rendering:
\begin{equation}
\int_{\Omega} \left[ -2(1-\phi)\delta \phi \, \mathcal{H} +
        G_c \left( C \right) \left( \dfrac{\phi}{\ell} \delta \phi
        + \ell\nabla \phi \cdot \nabla \delta \phi \right) \right]  \, \mathrm{d}V = 0  \, .
\end{equation}

Finally, the weak form for the hydrogen transport problem can be readily obtained by multiplying Eq. (\ref{eq:StrongCL}) by a test function $\delta C_L$ and integrating over the problem domain, such that:
\begin{equation}\label{Eq:WeakC}
    \int_\Omega \left[ \delta C_L \left( \frac{1}{D_e} \frac{dC_L}{dt} \right) + \nabla \delta C_L  \nabla C_L  - \nabla \delta C_L  \left( \frac{\bar{V}_H C_L }{RT}  \nabla \sigma_H \right) \right] \, \mathrm{d}V = 0 \, .
\end{equation}

Now make use of Voigt notation. The nodal variables for the displacement field $\mathbf{\hat{u}}$, the phase field $\hat{\phi}$ and the lattice hydrogen concentration $\hat{C}_L$ are interpolated as: 
\begin{equation}\label{eq:Ndiscret}
\mathbf{u} = \sum_{i=1}^m \bm{N}_i \hat{\mathbf{u}}_i, \hspace{1cm} \phi =  \sum_{i=1}^m N_i \hat{\phi}_i, \hspace{1cm} C_L =  \sum_{i=1}^m N_i \hat{C}_{L_i}.
\end{equation}
\noindent Here, $N_i$ denotes the shape function associated with node $i$ and $\bm{N}_i$ is the shape function matrix, a diagonal matrix with $N_i$ in the diagonal terms. Also, $m$ is the total number of nodes per element such that, assuming plane strain conditions, $\hat{\mathbf{u}}_i=\left\{ u_x, \, u_y \right\}^T$, $\hat{\phi}_i$ and $\hat{C}_{L_i}$ respectively denote the displacement, phase field and lattice hydrogen concentration at node $i$. Consequently, the associated gradient quantities can be discretised using the corresponding \textbf{B}-matrices, containing the derivative of the shape functions, such that:
\begin{equation}\label{eq:Bdiscret}
\bm{\varepsilon} = \sum\limits_{i=1}^m \bm{B}^{\bm{u}}_i \hat{\mathbf{u}}_i, \hspace{0.8cm}  \nabla\phi =  \sum\limits_{i=1}^m \mathbf{B}_i \hat{\phi}_i, \hspace{0.8cm} \nabla C_L =  \sum\limits_{i=1}^m \mathbf{B}_i \hat{C}_{L_i} \, .
\end{equation}

Considering the discretisation (\ref{eq:Ndiscret})-(\ref{eq:Bdiscret}), we derive the residuals for each primal kinematic variable as:
\begin{align}
    & \mathbf{R}_i^\mathbf{u} = \int_\Omega \left\{\left[\left(1-\phi\right)^2+ k\right]\left(\bm{B}^\mathbf{u}_i\right)^T \bm{\sigma}_0 \right\} \, \text{d}V \, , \\
    & R_i^\phi = \int_\Omega \left\{ -2\left(1-\phi\right)N_i \mathcal{H} + G_c (C) \left[\frac{\phi}{\ell}N_i + \ell \,  \left( \mathbf{B}_i \right)^T \nabla\phi\right]\right\}dV \, , \\
    & R_i^{C_L} = \int_{\Omega}\left[N_i\left(\dfrac{1}{D_e}\dfrac{\text{d}C_L}{\text{d}t}\right) + \mathbf{B}_i^T\nabla C_L - \mathbf{B}_i^T\left(\dfrac{\overline{V}_H C_L}{RT}\nabla\sigma_H\right)\right] \, \text{d}V  \, .
\end{align}

The consistent tangent stiffness matrices $\bm{K}$, required to complete the finite element implementation, are obtained by considering the constitutive relations and differentiating the residuals with respect to the incremental nodal variables as follows:
\begin{align}
    & \bm{K}_{ij}^{\mathbf{u}} = \frac{\partial \bm{R}_{i}^{\bm{u}} }{\partial \bm{u}_{j} } = 
        \int_{\Omega} \left\{ \left[ (1-\phi)^2+k \right] {(\bm{B}_{i}^{\bm{u}})}^{T} \bm{C} \, \bm{B}_{j}^{\bm{u}} \right\} \, \text{d}V \, , \\
    & \bm{K}_{ij}^{\phi} = \frac{\partial R_{i}^{\phi} }{ \partial \phi_{j} } =  \int_{\Omega} \left\{ \left( 2 \mathcal{H} + \frac{G_{c} \left( C \right)}{\ell} \right) N_{i} N_{j} + G_{c} \left( C \right) \ell \, \mathbf{B}_i^T\mathbf{B}_j \right\} \, \text{d}V \, , \\
    & \bm{K}^{C_L}_{ij} =\frac{\partial R^{C_L}_i}{\partial c_j} =\int_\Omega\left(N_i^T \dfrac{1}{D_e \text{d}t}N_j + \mathbf{B}_i^T\mathbf{B}_j - \mathbf{B}_i^T \frac{\overline{V}_H}{RT} \nabla \sigma_H N_j \right)\,\text{d}V \, .
\end{align}

The Newton-Raphson method is used to iteratively solve the global finite element system. A so-called staggered or alternative minimisation solution scheme is used, following (\citealp{Miehe2010a}; \citealp{CMAME2018}). A time sensitivity study is conducted in all computations.

\section{Results}
\label{Sec:FEMresults}

We proceed to demonstrate the potential of the theoretical and computational framework presented by simulating several boundary value problems of particular interest. First, in Section \ref{Sec:Rcurves}, the boundary layer concept is exploited to demonstrate that the model can capture the main experimental trends and to gain insight into the interplay between hydrogen and dislocation-hardening mechanisms. Then, in Section \ref{Sec:Experiments}, model predictions are benchmarked against experiments conducted on AISI 4135 steel. 

\subsection{Crack tip mechanics and growth resistance}
\label{Sec:Rcurves}

\begin{figure}[H]
\centering
\includegraphics[scale=1.5]{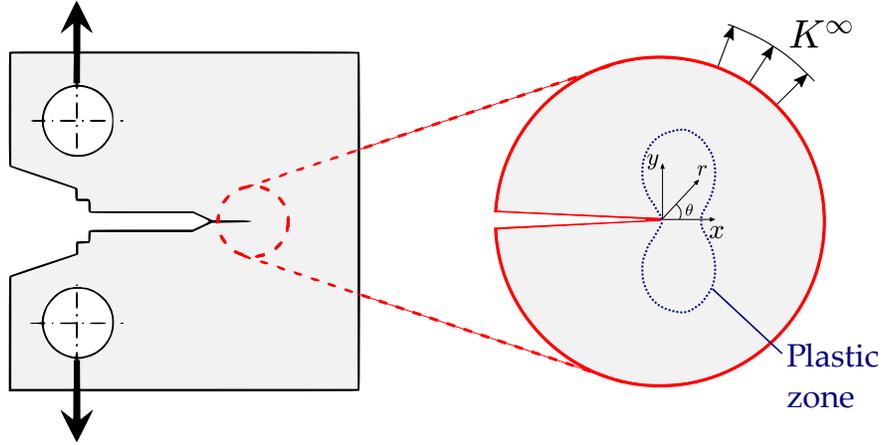}
\caption{Boundary layer concept under small scale yielding conditions, illustrated on a Compact Tension specimen.}
\label{fig:BoundaryLayer}
\end{figure}

We shall first investigate the mechanics of stationary and propagating cracks by considering the fracture of a metallic sample under small scale yielding conditions. The concept of a boundary layer formulation is illustrated in Fig. \ref{fig:BoundaryLayer}, using as example a Compact Tension specimen. For a cracked solid, the crack tip stress state is characterised by the stress intensity factor; $K_I$, assuming mode I conditions. The \citet{Williams1957} solution for a linear elastic solid can be used to relate the displacement field to the magnitude of $K_I$. Considering a polar coordinate system $(r, \theta)$ and a Cartesian coordinate system $(x, y)$ centred at the crack tip, with the crack plane along the negative $x$-axis, the displacement solution reads:
\begin{equation}\label{eq:Williams1}
u_i = \frac{K_I}{E} r^{1/2} f_i \left( \theta, \nu \right),
\end{equation}
\noindent where the subscript index $i$ equals $x$ or $y$, and the functions $f_i \left( \theta, \nu \right)$ are given by
\begin{equation}\label{eq:Williams2}
f_{x} = \frac{1+\nu}{\sqrt{2 \pi}} \left(3 - 4 \nu - \cos \theta \right) \, \cos \left(\frac{\theta}{2} \right)
\end{equation}
\begin{equation}\label{eq:Williams3}
f_{y} = \frac{1+\nu}{\sqrt{2 \pi}} \left(3 - 4 \nu - \cos \theta \right) \, \sin \left(\frac{\theta}{2} \right).
\end{equation}

After a mesh-sensitivity analysis, the finite element model is discretised with approximately 24,000 quadratic, quadrilateral elements with reduced integration. The mesh is uniformly refined in the vicinity of the crack and the region of crack extension, such that the fracture process is resolved by ensuring that the characteristic element length is at least 6 times smaller than the phase field length scale. In this boundary value problem, the results obtained are independent from the size of the crack and the radius of the specimen, as long as these two dimensions are significantly larger than the plastic zone size, $R_p$.

We simulate crack initiation and growth in a model steel with Young's modulus $E=200$ GPa, Poisson's ratio $\nu=0.3$, yield stress $\sigma_Y=600$ MPa, and strain hardening exponent $N=0.2$. Regarding the mass diffusion properties, a lattice diffusion coefficient of $D=0.0127$ mm$^2$/s is adopted, following \citet{Sofronis1989}. We consider the existence of three types of traps: dislocations, carbides and grain boundaries. The dislocation trap density $N_T^{(d)}$ is given by Eq. (\ref{Eq:NTrho}) and the binding energy is assumed to be $W_B^{(d)}=-20.2$ kJ/mol \citep{Hirth1980}. Carbide trap sites are characterised by a trap density of $\alpha N_T^{(c)}=8.464 \times 10^{17}$ sites/mm$^3$ \citep{Li2004} and a binding energy of $W_B^{(c)}=-11.5$ kJ/mol \citep{Dadfarnia2011}. Lastly, the trapping characteristics of grain boundaries are given by $W_B^{(gb)}=-30$ kJ/mol \citep{Serebrinsky2004} and $\alpha N_T^{(gb)}=8.464 \times 10^{13}$ sites/mm$^3$ \citep{Dadfarnia2011}.\\

To facilitate interpretation of the crack growth resistance results, we shall first investigate the behaviour of a stationary crack; i.e., disregarding the damage part of the formulation. Crack tip fields are shown normalising the distance ahead of the crack $r$ with Irwin's estimate of the plastic zone size:
\begin{equation}
    R_p = \frac{1}{3 \pi} \left( \frac{K_I}{\sigma_Y} \right)^2 \, .
\end{equation}

\begin{figure}[H]
\centering
\includegraphics[scale=0.65]{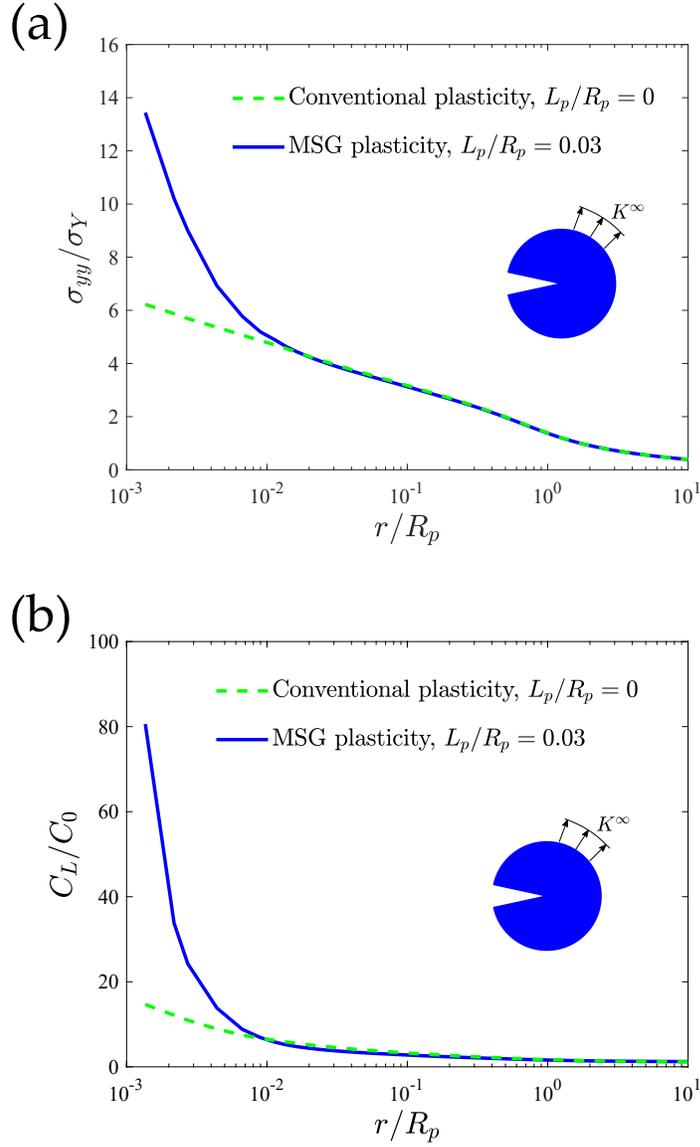}
\caption{Stationary crack tip analysis: (a) tensile stress and (b) lattice hydrogen concentration distributions ahead of the crack ($r,\theta=0^\circ$). Results are shown for both mechanism-based strain gradient (MSG) plasticity ($L_p/R_p=0.03$) and conventional plasticity ($L_p/R_p=0$).} 
\label{fig:S22CL}
\end{figure}

The distribution of the tensile stress component $\sigma_{yy}$ is shown in Fig. \ref{fig:S22CL}a for both the cases of $L_p=0.03R_p$ and $L_p=0$ (conventional plasticity). It is observed that dislocation hardening mechanisms associated with plastic strain gradients are negligible far away from the crack tip but become significant within a distance of $r/R_p=0.01$. The GND densities resulting from the large gradients of plastic strain present near the crack tip result in stress levels that are significantly larger than those predicted by conventional plasticity models ($L_p=0$). This is in agreement with expectations \citep{EJMAS2019}. The hydrostatic stress distribution shows similar differences between gradient-enriched and conventional plasticity predictions and this leads, in turn, to a larger hydrogen concentration. The lattice hydrogen concentration ahead of the crack tip is shown in Fig. \ref{fig:S22CL}b. The results have been obtained with a loading rate of $\dot{K}_I=0.183$ MPa$\sqrt{\text{m}}$/s and by defining an initial hydrogen concentration of $C_0=0.1$ wt ppm. The higher lattice hydrogen concentration predicted in the case of $L_p>0$ is not surprising given the dependence of the lattice hydrogen concentration on the hydrostatic stress; under steady state conditions, their relation reads:
\begin{equation}
   C_L = C_0 \exp \left( \frac{\bar{V}_H \sigma_H}{RT} \right) \, .
\end{equation}

\noindent Thus, small changes in the hydrostatic stress distribution can lead to significant differences in the lattice hydrogen concentration.

\begin{figure}[H]
\centering
\includegraphics[scale=0.75]{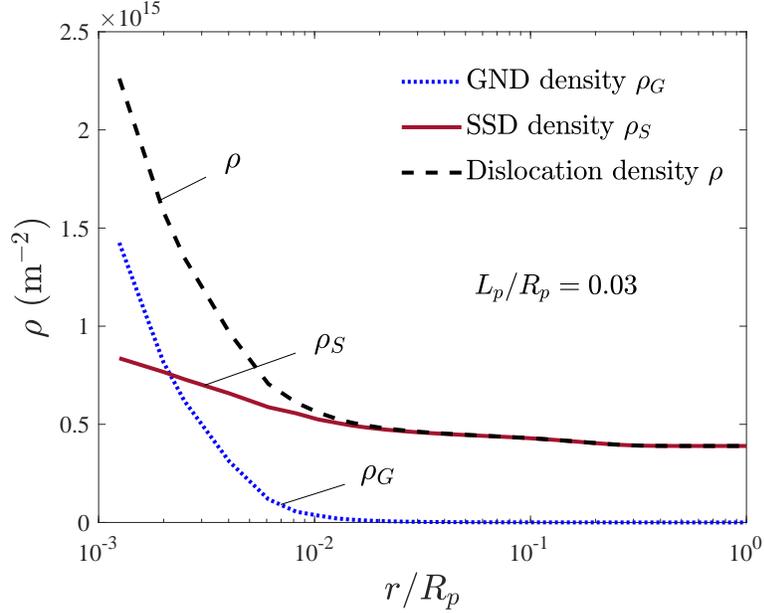}
\caption{Stationary crack tip analysis: dislocation density distribution (in m$^{-2}$) ahead of the crack ($r,\theta=0^\circ$). Results are shown for $L_p/R_p=0.03$, presenting the distribution of the GND dislocation density $\rho_G$, the SSD dislocation density $\rho_S$ and the total dislocation density $\rho$.}
\label{fig:rho}
\end{figure}

Moreover, the enriched crack tip mechanics description provided by the model captures another interesting effect. While local hardening due to strain gradients increases crack tip stresses and reduces the degree of plastic dissipation, this does not necessarily translate into a smaller crack tip dislocation density. The density of Statistically Stored Dislocations (SSDs) $\rho_S$ diminishes with increasing $L_p/R_p$ but this is counteracted by the associated increase in the density of Geometrically Necessary Dislocations (GNDs) $\rho_G$. The distributions of $\rho_G$ and $\rho_S$ are shown in Fig. \ref{fig:rho}. It can be observed that, for the choice $L_p/R_p=0.03$, the density of GNDs $\rho_G$ becomes larger than the density of SSDs $\rho_S$ close to the crack tip. This elevates the total density close to the crack tip, which has implications for the hydrogen trapped in dislocations.

\begin{figure}[H]
\centering
\includegraphics[scale=0.7]{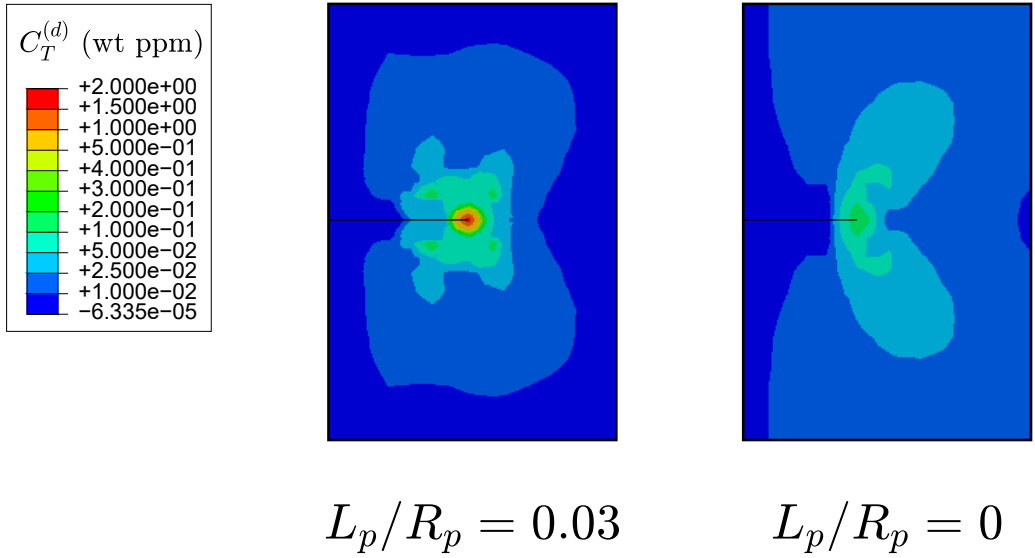}
\caption{Stationary crack tip analysis: Contours of the trapped hydrogen concentration at dislocation sites $C_T^{(d)}$. Results are shown for both mechanism-based strain gradient (MSG) plasticity ($L_p/R_p=0.03$) and conventional plasticity ($L_p/R_p=0$).}
\label{fig:CTD}
\end{figure}

The results shown in Fig. \ref{fig:CTD}, where the concentration of hydrogen trapped at dislocation sites $C_T^{(d)}$ is shown for both $L_p=0.03R_p$ and $L_p=0$, reveal that the consideration of mechanistic, strain gradient plasticity models can lead to a more significant effect of dislocation trap sites in the vicinity of the crack, relative to conventional plasticity models. For the binding energies and trap densities considered here, the hydrogen trapped at dislocation sites can reach levels of up to 2 wt ppm if the GND contribution to the dislocation trap density is accounted for. In all the coupled deformation-diffusion studies reported so far, the dislocation density is assumed to be that SSDs only, neglecting this important contribution (see, e.g., \citealp{Sofronis2001} and references therein). The higher $C_L$ levels predicted when $L_p>0$ are also likely contributing to the higher magnitude of $C_T^{(d)}$, as a higher trap occupancy is attained.  

Now, let us turn attention to crack propagation. The phase field fracture formulation for hydrogen embrittlement described in Section \ref{Sec:PhaseFieldTheory} is taken into consideration. Following (\citealp{Tvergaard1992}; \citealp{JMPS2019}), for a material with toughness $G_c$, a reference stress intensity fracture for crack initiation can be defined as,
\begin{equation}\label{eq:K0}
    K_0 = \left( \frac{E G_c}{1-\nu^2} \right)^{1/2} ,
\end{equation}

\noindent and a reference size of fracture process zone $R_0$ reads
\begin{equation}\label{eq:R0}
R_0 = \frac{1}{3 \pi \left( 1 - \nu^2 \right)} \frac{E G_c}{\sigma_Y^2} \, .   
\end{equation}

\noindent Thus, the relevant non-dimensional group for the phase field fracture process is given by $\ell/R_0$. As discussed extensively in \citet{Tanne2018}, 
the choice of a positive $\ell > 0^+$ in the phase field formulation introduces the concept of a material strength $\hat{\sigma}$; e.g., in a one-dimensional setting:
\begin{equation}\label{eq:Strength}
    \hat{\sigma} = \frac{9}{16} \sqrt{\frac{E G_c}{3 \ell}}
\end{equation}

\noindent Accordingly, the phase field length scale $\ell$ not only acts as a regularising parameter but is attributed a physical meaning. As discussed by \citet{JMPS2020}, an analogy can then be drawn with cohesive zone analyses based on material cohesive strength $\hat{\sigma}$. Specifically, combining (\ref{eq:R0}) and (\ref{eq:Strength}), one reaches,
\begin{equation}
    \frac{R_0}{\ell} = \frac{256}{81 \pi \left( 1 - \nu^2 \right)} \left( \frac{\hat{\sigma}}{\sigma_Y} \right)^2 \, .
\end{equation}

\begin{figure}[H]
\centering
\includegraphics[scale=0.7]{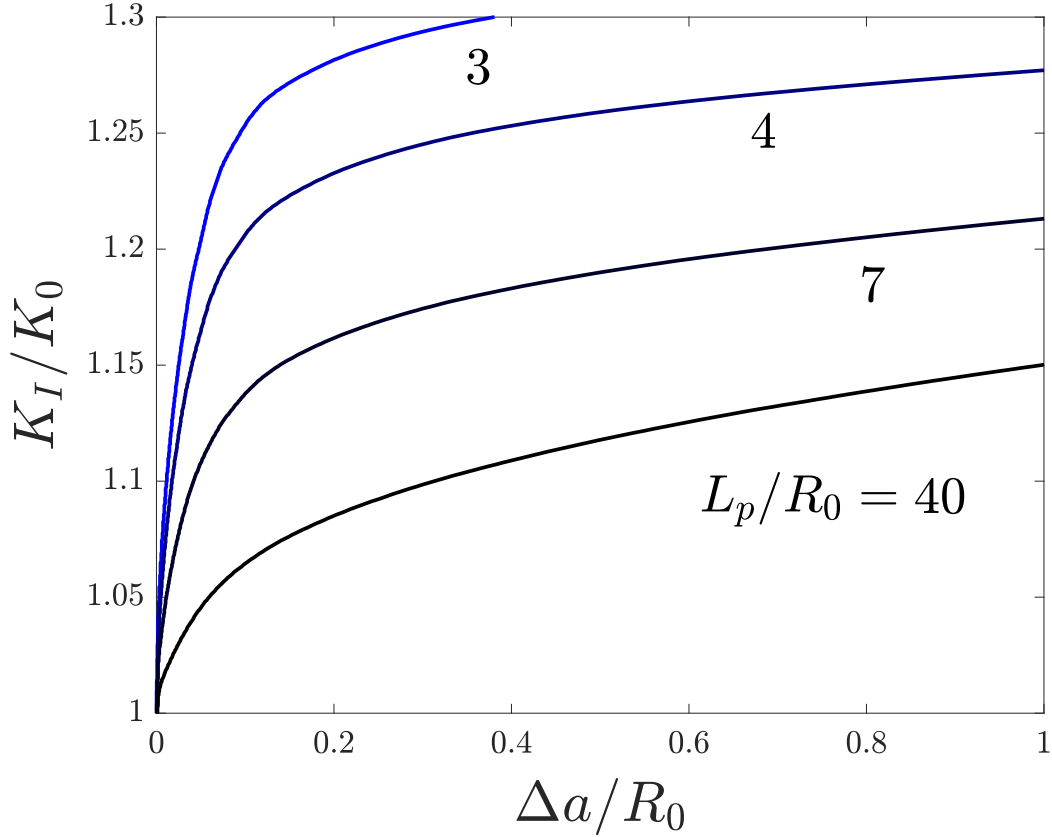}
\caption{Influence of the plastic length scale $(L_p)$  on crack growth resistance. Material properties: $\hat{\sigma}/\sigma_y \approx 6 \, (\ell / R_0 = 1/40)$, $\sigma_y / E =0.003$, $\nu=0.3$, $N=0.2$.}
\label{fig:L_p_effect}
\end{figure}

We shall first examine, in the absence of hydrogen, the role of the non-dimensional groups $L_p/R_0$ and $\ell/R_0$ (or $\hat{\sigma}/\sigma_Y$) in the fracture process. The results computed for the case of a varying $L_p/R_0$ are shown in Fig. \ref{fig:L_p_effect}. Results are presented in terms of a normalised remote load $K_I/K_0$ versus the normalised crack extension $\Delta a /R_0$. It can be observed that the initiation of crack growth takes place at $K_I = K_0$ (or $G=G_c$) for all cases, in agreement with expectations. However, the degree of dissipation is sensitive to the ratio between the plastic length scale and the fracture process zone. Larger $L_p/R_0$ values translate into a greater influence of plastic strain gradients and this results in a smaller fracture resistance. Quantitative predictions are thus sensitive to the capacity of the material to strengthen or harden in the presence of plastic strain gradients, characterised \textit{via} $L_p$, and on the work of fracture, as given by $G_c$. Also, this necessarily implies that gradient effects have a larger influence in brittle fracture processes, where $G_c$ is small and consequently $L_p/R_0$ is large. 

\begin{figure}[H]
\centering
\includegraphics[scale=0.63]{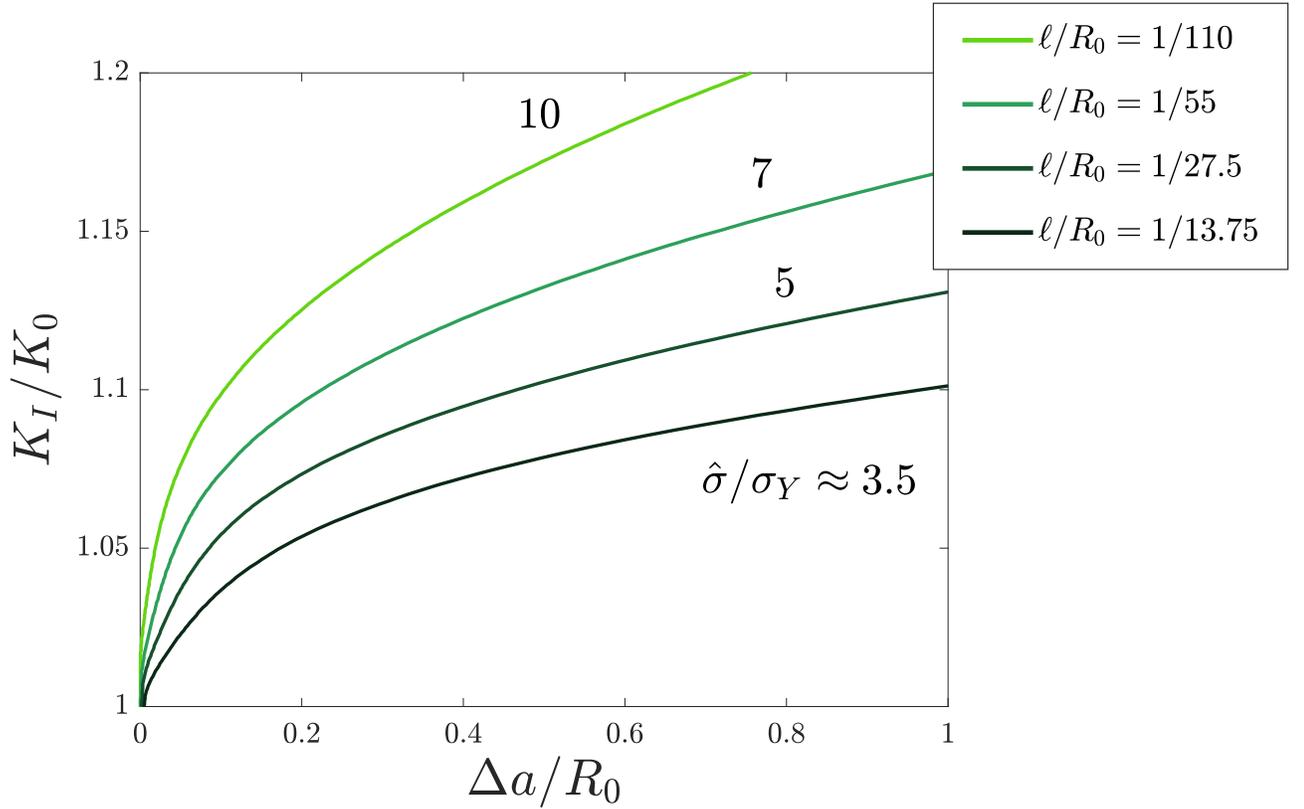}
\caption{Influence of the material strength $\hat{\sigma}/\sigma_Y$ ($\ell/R_0$) on crack growth resistance. Material properties: $\sigma_Y/E=0.003$, $\nu=0.3$, $N=0.2$, and $L_p/R_0=40$.} 
\label{fig:ell_effect}
\end{figure}

Consider now a fixed $L_p/R_0=40$ and vary the ratio of phase field length scale to fracture process zone size, $\ell/R_0$. We emphasise that this is equivalent to varying the material strength $\hat{\sigma}/\sigma_Y$. The results obtained are shown in Fig. \ref{fig:ell_effect}. Crack growth resistance diminishes with decreasing $\hat{\sigma}/\sigma_Y$ (increasing $\ell/R_0$). This agrees with the trends observed in cohesive zone model studies, where a higher crack growth resistance is observed for larger values of the cohesive strength. The larger the material strength the more plastic dissipation takes place during the crack propagation stage. It is important to note that brittle interfaces, such as grain boundaries, have strength values on the order of $\hat{\sigma}/\sigma_Y \sim 10$ and consequently for brittle fracture to be predicted two conditions must be met: (i) gradient effects must be considered, as otherwise the stress elevation is insufficient, and (ii) the magnitude of $G_c$ has to be small, e.g. through an embrittlement process, so that gradient effects ($L_p/R_0$) are sufficiently large. In the case of hydrogen embrittlement, there is a dual contribution of the hydrogen in reducing $G_c$ and the interface strength $\hat{\sigma}$, as shown in atomistic simulations (\citealp{VanderVen2003}; \citealp{Jiang2004a}; \citealp{Alvaro2015}), which enables predicting decohesion of brittle interfaces if GNDs and dislocation hardening are accounted for. In other words, the combination of strain gradient plasticity, a fracture process zone approach and a degradation of the toughness with hydrogen content provides a modern rationale for hydrogen enhanced decohesion.  

Now, let us explicitly incorporate the influence of hydrogen. For this, a constitutive choice for the hydrogen degradation law remains to be made. A mechanistic, multi-scale approach is adopted for this boundary value problem. The fracture process is assumed to be intergranular, as observed experimentally in many material systems exposed to hydrogen (see, e.g., \citealp{Banerji1978,Pouillier2012,Harris2018}), and thus driven by the hydrogen-assisted decohesion of grain boundaries - the occupancy of grain boundary trap sites $\theta^{(gb)}_T$ is the quantity of interest. A connection with the atomic scale process of grain boundary decohesion is established defining a degradation law that exhibits the linear decrease in surface energy with increasing hydrogen coverage observed in atomistic calculations \citep{Jiang2004a,Alvaro2015}:
\begin{equation}\label{eq:atomistic}
    G_c \left( \theta \right) = \left( 1 - \chi \theta^{(gb)}_T \right) G_c \left( 0 \right) \, .
\end{equation}

The parameter $\chi$ is a hydrogen damage coefficient, which can be fitted to quantitatively reproduce the atomistic results \citep{CMAME2018}. For example, here we adopt a value of $\chi=0.89$, which provides a good fit to the Density Functional Theory data for iron \citep{Jiang2004a}. The hydrogen transport properties are those described for the stationary crack analysis and the loading rate equals $\dot{K}_I/{K_0}=4\times10^{-7}$ s$^{-1}$. Samples are continuously exposed to an environmental hydrogen concentration $C_{env}$ and are exposed to the same environment for a sufficiently long time before mechanical loading is applied, such that $C_L=C_{env}$ $\forall \, x$ at $t=0$. A \emph{moving} chemical boundary condition is applied, as in \citep{CS2020}, to capture how the environment (hydrogen gas or aqueous electrolyte) promptly occupies the space created by crack advance.\\

\begin{figure}[H]
\centering
\includegraphics[scale=1]{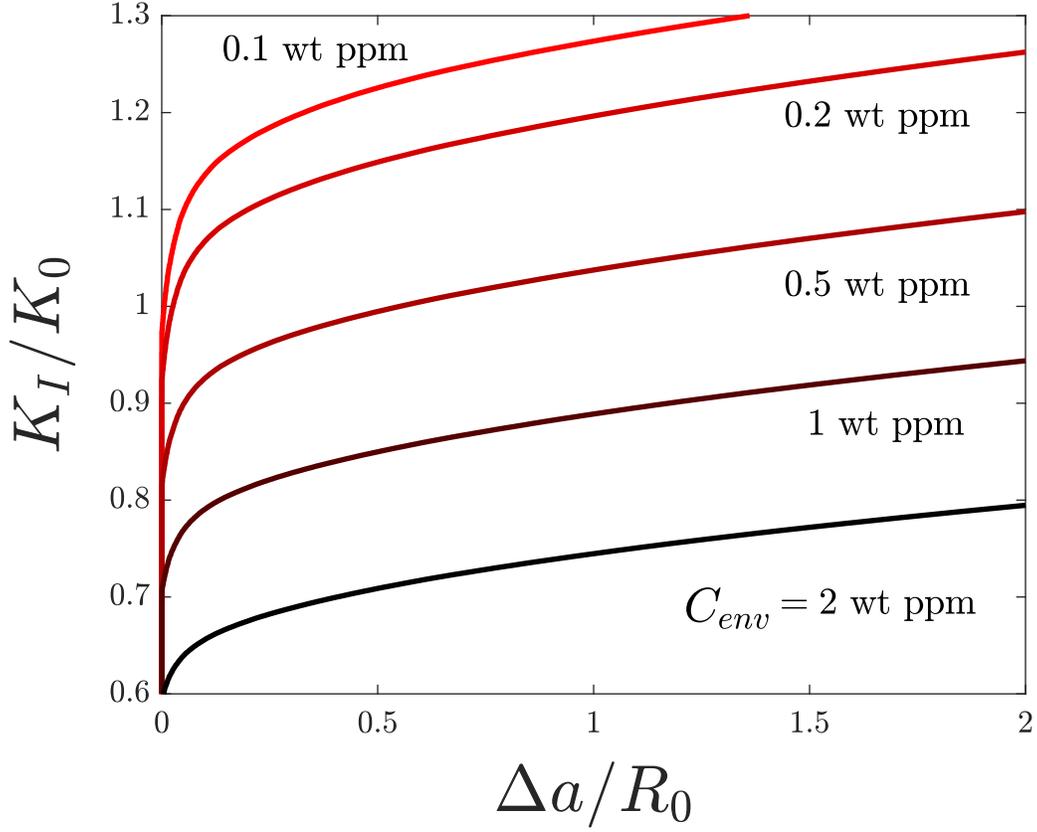}
\caption{Effect of the environmental hydrogen concentration on the fracture resistance.  Material properties: $\sigma_y/E=0.003$, $\nu=0.3$, $N=0.2$, $\hat{\sigma}/\sigma_y=10$ ($\ell/R_0=1/110$), $L_p/R_0=5$, $D=0.0127$ mm$^2$/s, and $\chi=0.89$. Loading rate $\dot{K}_I/{K_0}=4\times10^{-7}$ s$^{-1}$.}
\label{fig:H_Conc}
\end{figure}

The crack growth resistance curves obtained for different hydrogen environments are shown in Fig. \ref{fig:H_Conc}. The results show a significant decrease in the load for crack initiation with increasing hydrogen content; the initiation of crack growth takes place at $K_I$ values lower than hydrogen-free $K_0$ (computed from $G_c(0)$). Accordingly, the crack growth resistance also diminishes as the magnitude of $C_{env}$ increases. In addition, it is shown that the fracture of brittle interfaces ($\hat{\sigma}/\sigma_y=10$) can occur for hydrogen concentrations as low as 0.1 ppm if $L_p/R_0=5$.

Finally, we conclude the crack growth resistance analysis by assessing the influence of the loading rate. In this case, the specimens are pre-charged to a hydrogen content of 0.5 wt ppm but are then subjected to mechanical load in an inert environment, implying the application of a Dirichlet-type boundary condition $C_L=0$ in the crack surface. As in the hydrogen environmentally assisted cracking example, the $C_L=0$ condition is enforced at all times at the crack surfaces by means of a penalty approach. This reflects the fact that the newly created crack surfaces are immediately exposed to an inert environment. 

\begin{figure}[H]
\centering
\includegraphics[scale=0.7]{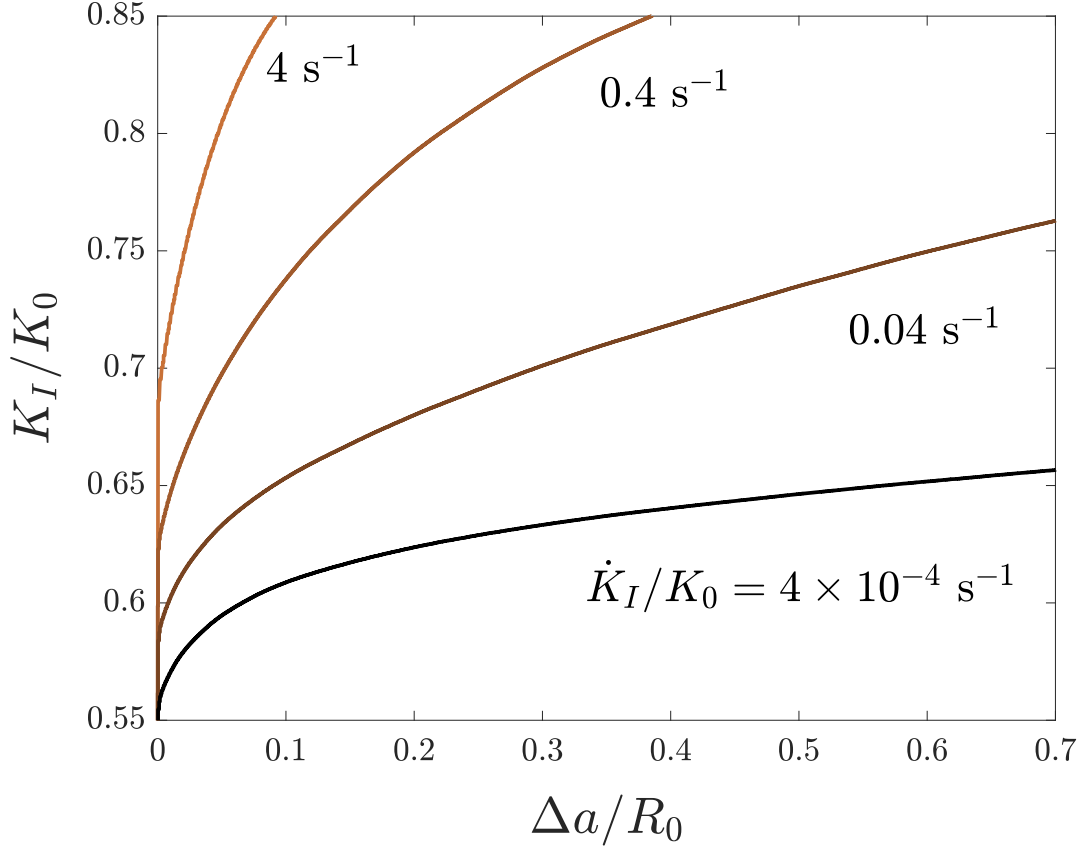}
\caption{Effect of the loading rate ($\dot{K}_I/K_0$) on the fracture resistance of samples pre-charged with a uniform hydrogen concentration of 0.5 wt ppm. Material properties: $\sigma_y/E=0.003$, $\nu=0.3$, $N=0.2$, $\hat{\sigma}/\sigma_y \approx 6$ ($\ell/R_0= 0.025$), $L_p/R_0=5$, $D=0.0127$ mm$^2$/s, and $\chi=0.89$.}
\label{fig:H_loading}
\end{figure}

The results obtained varying the loading rate from $\dot{K}_I/{K_0}=4\times10^{-7}$ to $\dot{K}_I/{K_0}=4$ s$^{-1}$ are shown in Fig. \ref{fig:H_loading}. The results reveal that the model can rigorously capture how hydrogen damage becomes more significant for smaller loading rates. The smaller the magnitude of $\dot{K}_I/K_0$, the larger the time available for the hydrogen to diffuse to the fracture process zone. The influence is observed over a range of loading rates spanning at least four orders of magnitude, consistent with experimental observations \citep{Momotani2017}.  

\subsection{Experimental validation: failure of pre-charged AISI 4135 steel bars}
\label{Sec:Experiments}

\begin{figure}[H]
\centering
\includegraphics[scale=1]{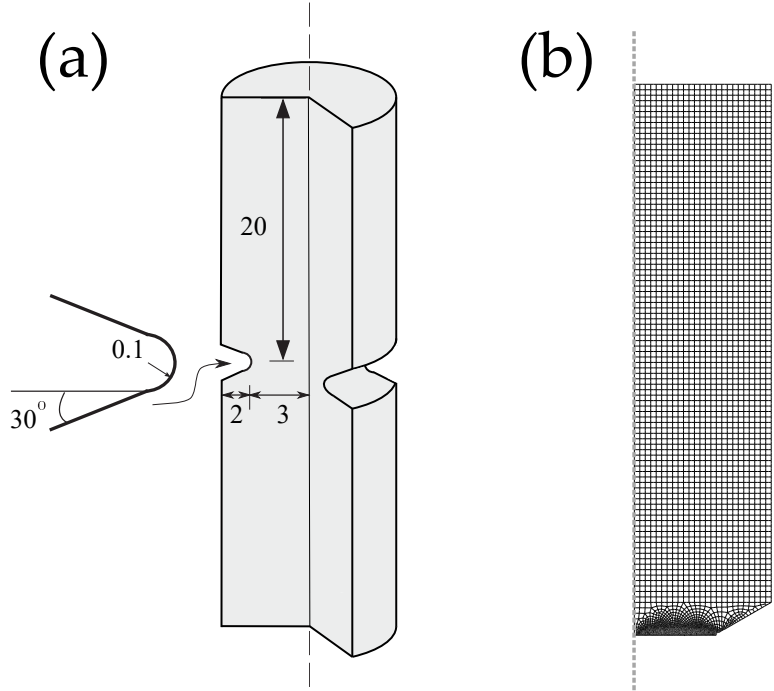}
\caption{Experimental comparison. Geometry, dimensions (in mm) and finite element mesh for the simulation of notch strength degradation of AISI 4135 steel bars with hydrogen content.}
\label{fig:ExperimentsSketch}
\end{figure}

We shall now compare model predictions with experimental measurements of failure stress versus hydrogen concentration. The experiments by \citet{Wang2005} on pre-charged notched AISI 4135 steel bars are taken as a benchmark. The geometry and dimensions are given in Fig. \ref{fig:ExperimentsSketch}. The bar is cylindrical and we thus adopt an axisymmetric formulation, modifying the strain-displacement matrix and integrating the discretised equations in polar coordinates. We also take advantage of symmetry and model only half of the plane problem. The model is discretised with a total of 20,373 quadratic quadrilateral axisymmetric elements with reduced integration. The finite element mesh is refined along the crack propagation region, with the characteristic element length being ten times smaller than the phase field length scale, $\ell=0.029$ mm. The material properties are listed in Table \ref{tab:tab1length}. Following \citet{Wang2005}, the mechanical properties of the AISI 4135 steel bars are given by a Young's modulus of $E=210$ GPa, a Poisson's ratio of $\nu=0.3$, and a yield stress of $\sigma_Y=1235$ MPa. The work hardening behaviour is captured by reproducing the stress-strain curve reported in \citep{Wang2005} with the hardening law given in Eq. (\ref{eq:PoweLaw}), this fitting exercise renders a strain hardening exponent of $N=0.05$.

The hydrogen transport properties are also listed in in Table \ref{tab:tab1length}. A lattice diffusion coefficient of $D=3.8 \times 10^{-11}$ m$^2$/s is considered, as reported in \citet{Wang2005}. Also, following their findings, two types of traps are considered: dislocations and grain boundaries. Quantitative values for the density and the binding energy of each trap type are not available. Thus, we take the trapping information from the study by \citet{AM2020} on AISI 4140 steel. Namely, the grain boundary trap density and binding energy are respectively given by $N_T^{(gb)}=5.06 \times 10^{25}$ sites/m$^3$ and $W_B^{(gb)}=-24.7$ kJ/mol. The dislocation trap binding energy is given by $W_B^{(d)}=-35.2$ kJ/mol, while $N_T^{(d)}$ is estimated from the total dislocation density by using Eq. (\ref{Eq:NTrho}). An initial dislocation density of $\rho_0=10^{14}$ m$^{-2}$ is assumed for the unstressed state. Fracture properties are not reported in the benchmark study by \citet{Wang2005}. Thus, the material toughness in the absence of hydrogen $G_c (0)$ is estimated by calibrating with the experiment conducted in air: the experimental notch tensile strength is attained with a magnitude of $G_c=25$ kJ/m$^2$.

\begin{table}[H]
\caption{Material properties used in the validation with the experiments on pre-charged AISI 4135 steel bars by \citet{Wang2005}. The parameters are taken from \citep{Wang2005} and \citep{AM2020}.}
\raggedleft
\hspace*{-3cm} 
\begin{tabular}{l l} 
\thickhline
Parameter & Magnitude\\
\thickhline
Mechanical properties\\
\hline
Young's modulus, $E$ & 210,000 MPa \\
Poisson's ratio, $\nu$ & 0.3\\
Yield stress, $\sigma_Y$ & 1235 MPa \\
Strain hardening exponent, $N$ & 0.05  \\
\hline
Hydrogen transport properties\\
\hline
Lattice diffusion coefficient, $D$ & 3.8$\times$10$^{-11}$ m$^2$/s \\
Grain boundary trap density, $N_T^{(gb)}$ & 5.06$\times$10$^{25}$ sites/m$^3$ \\
Grain boundary binding energy, $W_B^{(gb)}$ & -24.7 kJ/mol \\
Initial dislocation trap density, $N_T^{(d)}(t=0)$ & 5.06$\times$10$^{25}$ sites/m$^3$ \\
Dislocation binding energy, $W_B^{(d)}$ & -35.2 kJ/mol \\
\thickhline
\end{tabular}
\label{tab:tab1length}
\end{table}

To capture the hydrogen degradation, a phenomenological approach is followed, building upon the experimental data available: sensitivity of the notch tensile strength $\sigma_f$ versus pre-charged hydrogen content $C_0$. First, we conduct a series of virtual experiments in the absence of hydrogen to determine the magnitude of $G_c$ that corresponds to each critical stress value. We then estimate the magnitude of $\theta_T^{(gb)}$ that corresponds to the value of the lattice hydrogen concentration at the beginning of the experiment. Thus, assuming Oriani's equilibrium, the occupancy of grain boundary trapping sites can be determined for a given $C_L$ by considering Eqs. (\ref{eq:Oriani}) and (\ref{eq:KT}). This allows plotting $G_c$ vs $\theta_T^{(gb)}$, which is accurately fitted with the following law:
\begin{equation}\label{eq:phenomenological}
    G_c = G_c \left( 0 \right) \left( 3.519 \exp (-103.1\theta_T^{(gb)}) + 0.1567\exp (-5.572\theta_T^{(gb)}) \right)
\end{equation}

\begin{figure}[H]
\centering
\includegraphics[scale=1]{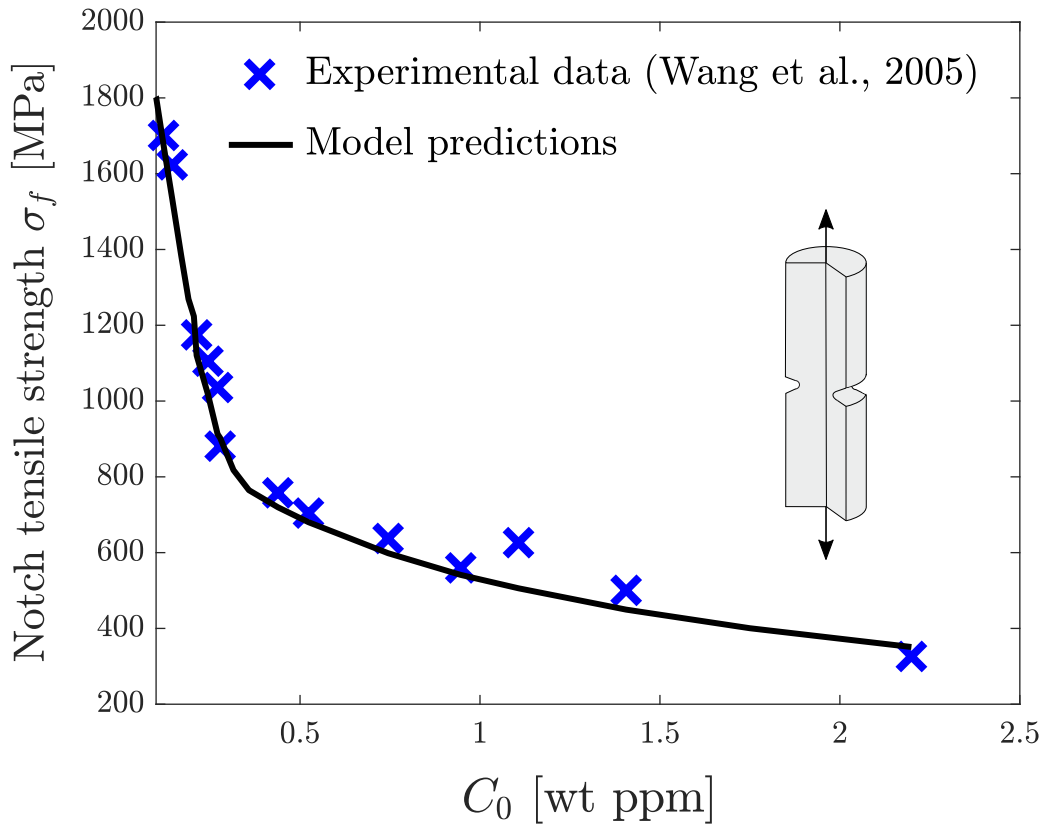}
\caption{Experimental comparison. Predictions of the reduction in tensile strength with pre-charged hydrogen concentration for notched AISI 4135 steel bars. The solid lines denote model predictions (present work) while the symbols correspond to the experimental data reported by \citet{Wang2005}.}
\label{fig:Experiments}
\end{figure}

This two-term exponential relation that provides the best fit to the simulated $G_c$ vs $\theta_T^{(gb)}$ data is introduced into the model, replacing the atomistic law employed in Section \ref{Sec:Rcurves}. We proceed then to run a number of deformation-diffusion-fracture computations to determine the notch failure strength as a function of the pre-charged hydrogen content. The strain loading rate equals $\dot{\varepsilon}=8.3 \times 10^{-7}$ s$^{-1}$, as in the experiments by \citet{Wang2005}. The results obtained are shown in Fig. \ref{fig:Experiments}, together with the experimental data.

An excellent agreement with the experimental results by \citet{Wang2005} is obtained. This is not surprising, given the phenomenological approach adopted, where the experimental data is used to construct the degradation law. It is of interest to compare this degradation law with the atomistic one adopted in Section \ref{Sec:Rcurves} - Eq. (\ref{eq:atomistic}). Values of $\theta_T^{(gb)}$ equal or higher than 0.2 result in a significantly larger degradation for the phenomenological approach. In fact, Eq. (\ref{eq:phenomenological}) exhibits a sharp drop for small values of $\theta_T^{(gb)}$ ($\leq 0.05$), followed by a milder slope (relative to the atomistic curve). The phenomenological degradation law is a reflection of the experimental results; as shown (e.g.) in Fig. 1 of \citep{AM2020}, Oriani's equilibrium dictates that for a trap with binding energy $W_B^{(gb)}=-24.7$ kJ/mol, an occupancy of $\theta_T=0.05$ is attained for $C_L \approx 0.4$ ppm. Thus, experiments suggest a reduction in notch tensile strength of more than 50\% for $\theta_T^{(gb)}=0.05$. Is such a small occupancy value capable of bringing such a significant reduction in grain boundary strength or are there other mechanisms involved? We note that this analysis is based on the bulk concentration; while this is reasonable for notched samples, the consideration of a small pre-existing crack would lead to much higher local levels of $C_L$ and grain boundary trap occupancy. Also, these estimations are very sensitive to the magnitude of $W_B^{(gb)}$. If the trap binding energy of -35.2 kJ/mol were to be interpreted as that of grain boundaries (as opposed to dislocations), the grain boundary occupancy would be approximately $\theta_T^{(gb)}=0.7$ for $C_L = 0.4$ ppm, bringing the phenomenological and atomistic laws very close to each other. Undoubtedly, there is a strong need to develop rigorous procedures for quantifying the hydrogen trapping characteristics of metals.

\section{Conclusions}
\label{Sec:ConcludingRemarks}

We have presented a new mechanistic framework for predicting the embrittlement of metallic components exposed to hydrogen-containing environments. Key features of the model include: (i) a mechanism-based strain gradient constitutive characterisation of crack tip stresses and dislocation densities, (ii) a coupled deformation-diffusion transport formulation, accounting for multiple trap types, and (iii) a hydrogen-dependent phase field description of fracture. The model was numerically implemented in the context of the finite element method, with displacements, hydrogen concentration and phase field parameter being the primary kinematic variables. First, stationary cracks were investigated to assess the influence on crack tip mechanics of the Taylor-based constitutive model adopted. Results showed that large plastic strain gradients close to the crack led to crack tip stresses that were notably higher than those predicted using conventional continuum models. We also investigated the interplay between the various length scales involved: the fracture process zone length $R_0$, the plastic length scale $L_p$ and the phase field length scale $\ell$, which governs the material strength. Also, it was shown that the crack growth resistance decreases with increasing strength, as there is a greater degree of plastic dissipation, and with $L_p/R_0$, as gradient effects become more significant. Thirdly, model predictions were benchmarked against experiments on cylindrical notched bars made of AISI 4135 steel, which were pre-charged with different levels of hydrogen content. The results showed that the model can quantitatively capture how the notch tensile strength drops with increasing hydrogen concentration. During the analyses, two options were considered to define the degradation of the material toughness with hydrogen: (i) a \emph{first principles} approach, in which atomistic calculations are used to establish the relation between fracture energy and hydrogen coverage, and (ii) a new phenomenological procedure, which is based on experimental data but does not require conducting coupled simulations. It was shown that both approaches led to a similar sensitivity of the fracture energy to hydrogen coverage if grain boundary cracking drives embrittlement and if the grain boundary trap binding energy lies within the range of -35 to -40 kJ/mol. These results highlight the need for an accurate characterisation of the hydrogen diffusion energy landscape.

\section{Acknowledgements}
\label{Sec:Acknowledgeoffunding}

The authors acknowledge financial support from the UK Engineering and Physical Sciences Research Council (EPSRC) through grants EP/R513052/1 and EP/V04902X/1.







\bibliographystyle{elsarticle-harv}
\bibliography{library}

\begin{thebibliography}{94}
\expandafter\ifx\csname natexlab\endcsname\relax\def\natexlab#1{#1}\fi
\providecommand{\url}[1]{\texttt{#1}}
\providecommand{\href}[2]{#2}
\providecommand{\path}[1]{#1}
\providecommand{\DOIprefix}{doi:}
\providecommand{\ArXivprefix}{arXiv:}
\providecommand{\URLprefix}{URL: }
\providecommand{\Pubmedprefix}{pmid:}
\providecommand{\doi}[1]{\href{http://dx.doi.org/#1}{\path{#1}}}
\providecommand{\Pubmed}[1]{\href{pmid:#1}{\path{#1}}}
\providecommand{\bibinfo}[2]{#2}
\ifx\xfnm\relax \def\xfnm[#1]{\unskip,\space#1}\fi
\bibitem[{Alvaro et~al.(2015)Alvaro, {Thue Jensen}, Kheradmand, L{\o}vvik and
  Olden}]{Alvaro2015}
\bibinfo{author}{Alvaro, A.}, \bibinfo{author}{{Thue Jensen}, I.},
  \bibinfo{author}{Kheradmand, N.}, \bibinfo{author}{L{\o}vvik, O.M.},
  \bibinfo{author}{Olden, V.}, \bibinfo{year}{2015}.
\newblock \bibinfo{title}{{Hydrogen embrittlement in nickel, visited by first
  principles modeling, cohesive zone simulation and nanomechanical testing}}.
\newblock \bibinfo{journal}{International Journal of Hydrogen Energy}
  \bibinfo{volume}{40}, \bibinfo{pages}{16892--16900}.
\bibitem[{Ambati et~al.(2015)Ambati, Gerasimov and {De Lorenzis}}]{Ambati2015}
\bibinfo{author}{Ambati, M.}, \bibinfo{author}{Gerasimov, T.},
  \bibinfo{author}{{De Lorenzis}, L.}, \bibinfo{year}{2015}.
\newblock \bibinfo{title}{{A review on phase-field models of brittle fracture
  and a new fast hybrid formulation}}.
\newblock \bibinfo{journal}{Computational Mechanics} \bibinfo{volume}{55},
  \bibinfo{pages}{383--405}.
\bibitem[{Amor et~al.(2009)Amor, Marigo and Maurini}]{Amor2009}
\bibinfo{author}{Amor, H.}, \bibinfo{author}{Marigo, J.J.},
  \bibinfo{author}{Maurini, C.}, \bibinfo{year}{2009}.
\newblock \bibinfo{title}{{Regularized formulation of the variational brittle
  fracture with unilateral contact: Numerical experiments}}.
\newblock \bibinfo{journal}{Journal of the Mechanics and Physics of Solids}
  \bibinfo{volume}{57}, \bibinfo{pages}{1209--1229}.
\bibitem[{Anand et~al.(2019)Anand, Mao and Talamini}]{Anand2019}
\bibinfo{author}{Anand, L.}, \bibinfo{author}{Mao, Y.},
  \bibinfo{author}{Talamini, B.}, \bibinfo{year}{2019}.
\newblock \bibinfo{title}{{On modeling fracture of ferritic steels due to
  hydrogen embrittlement}}.
\newblock \bibinfo{journal}{Journal of the Mechanics and Physics of Solids}
  \bibinfo{volume}{122}, \bibinfo{pages}{280--314}.
\bibitem[{Arsenlis and Parks(1999)}]{Arsenlis1999}
\bibinfo{author}{Arsenlis, A.}, \bibinfo{author}{Parks, D.M.},
  \bibinfo{year}{1999}.
\newblock \bibinfo{title}{{Crystallographic aspects of geometrically-necessary
  and statistically-stored dislocation density}}.
\newblock \bibinfo{journal}{Acta Materialia} \bibinfo{volume}{47},
  \bibinfo{pages}{1597--1611}.
\bibitem[{Banerji et~al.(1978)Banerji, McMahon and Feng}]{Banerji1978}
\bibinfo{author}{Banerji, S.K.}, \bibinfo{author}{McMahon, C.J.},
  \bibinfo{author}{Feng, H.C.}, \bibinfo{year}{1978}.
\newblock \bibinfo{title}{{Intergranular fracture in 4340-type steels: Effects
  of impurities and hydrogen}}.
\newblock \bibinfo{journal}{Metallurgical Transactions A} \bibinfo{volume}{9},
  \bibinfo{pages}{237--247}.
\bibitem[{Barrera et~al.(2016)Barrera, Tarleton, Tang and Cocks}]{Barrera2016}
\bibinfo{author}{Barrera, O.}, \bibinfo{author}{Tarleton, E.},
  \bibinfo{author}{Tang, H.W.}, \bibinfo{author}{Cocks, A.C.F.},
  \bibinfo{year}{2016}.
\newblock \bibinfo{title}{{Modelling the coupling between hydrogen diffusion
  and the mechanical behaviour of metals}}.
\newblock \bibinfo{journal}{Computational Materials Science}
  \bibinfo{volume}{122}, \bibinfo{pages}{219--228}.
\bibitem[{Bellettini and Coscia(1994)}]{Bellettini1994}
\bibinfo{author}{Bellettini, G.}, \bibinfo{author}{Coscia, A.},
  \bibinfo{year}{1994}.
\newblock \bibinfo{title}{{Discrete approximation of a free discontinuity
  problem}}.
\newblock \bibinfo{journal}{Numerical Functional Analysis and Optimization}
  \bibinfo{volume}{15}, \bibinfo{pages}{201--224}.
\bibitem[{Borden et~al.(2016)Borden, Hughes, Landis, Anvari and
  Lee}]{Borden2016}
\bibinfo{author}{Borden, M.J.}, \bibinfo{author}{Hughes, T.J.R.},
  \bibinfo{author}{Landis, C.M.}, \bibinfo{author}{Anvari, A.},
  \bibinfo{author}{Lee, I.J.}, \bibinfo{year}{2016}.
\newblock \bibinfo{title}{{A phase-field formulation for fracture in ductile
  materials: Finite deformation balance law derivation, plastic degradation,
  and stress triaxiality effects}}.
\newblock \bibinfo{journal}{Computer Methods in Applied Mechanics and
  Engineering} \bibinfo{volume}{312}, \bibinfo{pages}{130--166}.
\bibitem[{Bourdin et~al.(2000)Bourdin, Francfort and Marigo}]{Bourdin2000}
\bibinfo{author}{Bourdin, B.}, \bibinfo{author}{Francfort, G.A.},
  \bibinfo{author}{Marigo, J.J.}, \bibinfo{year}{2000}.
\newblock \bibinfo{title}{{Numerical experiments in revisited brittle
  fracture}}.
\newblock \bibinfo{journal}{Journal of the Mechanics and Physics of Solids}
  \bibinfo{volume}{48}, \bibinfo{pages}{797--826}.
\bibitem[{del Busto et~al.(2017)del Busto, Beteg{\'{o}}n and
  Mart{\'{i}}nez-Pa{\~{n}}eda}]{EFM2017}
\bibinfo{author}{del Busto, S.}, \bibinfo{author}{Beteg{\'{o}}n, C.},
  \bibinfo{author}{Mart{\'{i}}nez-Pa{\~{n}}eda, E.}, \bibinfo{year}{2017}.
\newblock \bibinfo{title}{{A cohesive zone framework for environmentally
  assisted fatigue}}.
\newblock \bibinfo{journal}{Engineering Fracture Mechanics}
  \bibinfo{volume}{185}, \bibinfo{pages}{210--226}.
\bibitem[{Chambolle(2004)}]{Chambolle2004}
\bibinfo{author}{Chambolle, A.}, \bibinfo{year}{2004}.
\newblock \bibinfo{title}{{An approximation result for special functions with
  bounded deformation}}.
\newblock \bibinfo{journal}{Journal des Mathematiques Pures et Appliquees}
  \bibinfo{volume}{83}, \bibinfo{pages}{929--954}.
\bibitem[{Cui et~al.(2021)Cui, Ma and Mart{\'{i}}nez-Pa{\~{n}}eda}]{JMPS2021}
\bibinfo{author}{Cui, C.}, \bibinfo{author}{Ma, R.},
  \bibinfo{author}{Mart{\'{i}}nez-Pa{\~{n}}eda, E.}, \bibinfo{year}{2021}.
\newblock \bibinfo{title}{{A phase field formulation for dissolution-driven
  stress corrosion cracking}}.
\newblock \bibinfo{journal}{Journal of the Mechanics and Physics of Solids}
  \bibinfo{volume}{147}, \bibinfo{pages}{104254}.
\bibitem[{Dadfarnia et~al.(2010)Dadfarnia, Novak, Ahn, Liu, Sofronis, Johnson
  and Robertson}]{Dadfarnia2010}
\bibinfo{author}{Dadfarnia, M.}, \bibinfo{author}{Novak, P.},
  \bibinfo{author}{Ahn, D.C.}, \bibinfo{author}{Liu, J.B.},
  \bibinfo{author}{Sofronis, P.}, \bibinfo{author}{Johnson, D.D.},
  \bibinfo{author}{Robertson, I.M.}, \bibinfo{year}{2010}.
\newblock \bibinfo{title}{{Recent advances in the study of structural materials
  compatibility with hydrogen}}.
\newblock \bibinfo{journal}{Advanced Materials} \bibinfo{volume}{22},
  \bibinfo{pages}{1128--1135}.
\bibitem[{Dadfarnia et~al.(2011)Dadfarnia, Sofronis and Neeraj}]{Dadfarnia2011}
\bibinfo{author}{Dadfarnia, M.}, \bibinfo{author}{Sofronis, P.},
  \bibinfo{author}{Neeraj, T.}, \bibinfo{year}{2011}.
\newblock \bibinfo{title}{{Hydrogen interaction with multiple traps: Can it be
  used to mitigate embrittlement?}}
\newblock \bibinfo{journal}{International Journal of Hydrogen Energy}
  \bibinfo{volume}{36}, \bibinfo{pages}{10141--10148}.
\bibitem[{Davey(1925)}]{Davey1925}
\bibinfo{author}{Davey, W.P.}, \bibinfo{year}{1925}.
\newblock \bibinfo{title}{{Precision measurements of the lattice constants of
  twelve common metals}}.
\newblock \bibinfo{journal}{Physical Review} \bibinfo{volume}{25},
  \bibinfo{pages}{753--761}.
\bibitem[{{Di Leo} and Anand(2013)}]{DiLeo2013}
\bibinfo{author}{{Di Leo}, C.V.}, \bibinfo{author}{Anand, L.},
  \bibinfo{year}{2013}.
\newblock \bibinfo{title}{{Hydrogen in metals: A coupled theory for species
  diffusion and large elastic-plastic deformations}}.
\newblock \bibinfo{journal}{International Journal of Plasticity}
  \bibinfo{volume}{43}, \bibinfo{pages}{42--69}.
\bibitem[{D{\'{i}}az et~al.(2016)D{\'{i}}az, Alegre and Cuesta}]{Diaz2016b}
\bibinfo{author}{D{\'{i}}az, A.}, \bibinfo{author}{Alegre, J.M.},
  \bibinfo{author}{Cuesta, I.I.}, \bibinfo{year}{2016}.
\newblock \bibinfo{title}{{Coupled hydrogen diffusion simulation using a heat
  transfer analogy}}.
\newblock \bibinfo{journal}{International Journal of Mechanical Sciences}
  \bibinfo{volume}{115-116}, \bibinfo{pages}{360--369}.
\bibitem[{Djukic et~al.(2019)Djukic, Bakic, {Sijacki Zeravcic}, Sedmak and
  Rajicic}]{Djukic2019}
\bibinfo{author}{Djukic, M.B.}, \bibinfo{author}{Bakic, G.M.},
  \bibinfo{author}{{Sijacki Zeravcic}, V.}, \bibinfo{author}{Sedmak, A.},
  \bibinfo{author}{Rajicic, B.}, \bibinfo{year}{2019}.
\newblock \bibinfo{title}{{The synergistic action and interplay of hydrogen
  embrittlement mechanisms in steels and iron: Localized plasticity and
  decohesion}}.
\newblock \bibinfo{journal}{Engineering Fracture Mechanics}
  \bibinfo{volume}{216}, \bibinfo{pages}{106528}.
\bibitem[{Duda et~al.(2015)Duda, Ciarbonetti, S{\'{a}}nchez and
  Huespe}]{Duda2015}
\bibinfo{author}{Duda, F.P.}, \bibinfo{author}{Ciarbonetti, A.},
  \bibinfo{author}{S{\'{a}}nchez, P.J.}, \bibinfo{author}{Huespe, A.E.},
  \bibinfo{year}{2015}.
\newblock \bibinfo{title}{{A phase-field/gradient damage model for brittle
  fracture in elastic-plastic solids}}.
\newblock \bibinfo{journal}{International Journal of Plasticity}
  \bibinfo{volume}{65}, \bibinfo{pages}{269--296}.
\bibitem[{Duda et~al.(2018)Duda, Ciarbonetti, Toro and Huespe}]{Duda2018}
\bibinfo{author}{Duda, F.P.}, \bibinfo{author}{Ciarbonetti, A.},
  \bibinfo{author}{Toro, S.}, \bibinfo{author}{Huespe, A.E.},
  \bibinfo{year}{2018}.
\newblock \bibinfo{title}{{A phase-field model for solute-assisted brittle
  fracture in elastic-plastic solids}}.
\newblock \bibinfo{journal}{International Journal of Plasticity}
  \bibinfo{volume}{102}, \bibinfo{pages}{16--40}.
\bibitem[{Elmukashfi et~al.(2020)Elmukashfi, Tarleton and
  Cocks}]{Elmukashfi2020}
\bibinfo{author}{Elmukashfi, E.}, \bibinfo{author}{Tarleton, E.},
  \bibinfo{author}{Cocks, A.C.F.}, \bibinfo{year}{2020}.
\newblock \bibinfo{title}{{A modelling framework for coupled hydrogen diffusion
  and mechanical behaviour of engineering components}}.
\newblock \bibinfo{journal}{Computational Mechanics} \bibinfo{volume}{66},
  \bibinfo{pages}{189--220}.
\bibitem[{Fern{\'{a}}ndez-Sousa et~al.(2020)Fern{\'{a}}ndez-Sousa,
  Beteg{\'{o}}n and Mart{\'{i}}nez-Pa{\~{n}}eda}]{AM2020}
\bibinfo{author}{Fern{\'{a}}ndez-Sousa, R.}, \bibinfo{author}{Beteg{\'{o}}n,
  C.}, \bibinfo{author}{Mart{\'{i}}nez-Pa{\~{n}}eda, E.}, \bibinfo{year}{2020}.
\newblock \bibinfo{title}{{Analysis of the influence of microstructural traps
  on hydrogen assisted fatigue}}.
\newblock \bibinfo{journal}{Acta Materialia} \bibinfo{volume}{199},
  \bibinfo{pages}{253--263}.
\bibitem[{Fleck and Hutchinson(1997)}]{Fleck1997}
\bibinfo{author}{Fleck, N.A.}, \bibinfo{author}{Hutchinson, J.W.},
  \bibinfo{year}{1997}.
\newblock \bibinfo{title}{{Strain gradient plasticity}}.
\newblock \bibinfo{journal}{Advances in Applied Mechanics}
  \bibinfo{volume}{33}, \bibinfo{pages}{295--361}.
\bibitem[{Fleck et~al.(1994)Fleck, Muller, Ashby and Hutchinson}]{Fleck1994}
\bibinfo{author}{Fleck, N.A.}, \bibinfo{author}{Muller, G.M.},
  \bibinfo{author}{Ashby, M.F.}, \bibinfo{author}{Hutchinson, J.W.},
  \bibinfo{year}{1994}.
\newblock \bibinfo{title}{{Strain gradient plasticity: Theory and Experiment}}.
\newblock \bibinfo{journal}{Acta Metallurgica et Materialia}
  \bibinfo{volume}{42}, \bibinfo{pages}{475--487}.
\bibitem[{Gangloff(2003)}]{Gangloff2003}
\bibinfo{author}{Gangloff, R.P.}, \bibinfo{year}{2003}.
\newblock \bibinfo{title}{{Hydrogen-assisted Cracking}}, in:
  \bibinfo{editor}{Milne, I.}, \bibinfo{editor}{Ritchie, R.},
  \bibinfo{editor}{Karihaloo, B.} (Eds.), \bibinfo{booktitle}{Comprehensive
  Structural Integrity Vol. 6}. \bibinfo{publisher}{Elsevier Science},
  \bibinfo{address}{New York, NY}, pp. \bibinfo{pages}{31--101}.
\bibitem[{Gangloff and Somerday(2012)}]{Gangloff2012}
\bibinfo{author}{Gangloff, R.P.}, \bibinfo{author}{Somerday, B.P.},
  \bibinfo{year}{2012}.
\newblock \bibinfo{title}{{Gaseous Hydrogen Embrittlement of Materials in
  Energy Technologies}}.
\newblock \bibinfo{publisher}{Woodhead Publishing Limited},
  \bibinfo{address}{Cambridge}.
\bibitem[{Gao et~al.(1999)Gao, Hang, Nix and Hutchinson}]{Gao1999}
\bibinfo{author}{Gao, H.}, \bibinfo{author}{Hang, Y.}, \bibinfo{author}{Nix,
  W.D.}, \bibinfo{author}{Hutchinson, J.W.}, \bibinfo{year}{1999}.
\newblock \bibinfo{title}{{Mechanism-based strain gradient plasticity - I.
  Theory}}.
\newblock \bibinfo{journal}{Journal of the Mechanics and Physics of Solids}
  \bibinfo{volume}{47}, \bibinfo{pages}{1239--1263}.
\bibitem[{Gurtin and Anand(2005)}]{Gurtin2005a}
\bibinfo{author}{Gurtin, M.E.}, \bibinfo{author}{Anand, L.},
  \bibinfo{year}{2005}.
\newblock \bibinfo{title}{{A theory of strain-gradient plasticity for
  isotropic, plastically irrotational materials. Part II: Finite
  deformations}}.
\newblock \bibinfo{journal}{International Journal of Plasticity}
  \bibinfo{volume}{21}, \bibinfo{pages}{2297--2318}.
\bibitem[{Gurtin et~al.(2010)Gurtin, Fried and Anand}]{Gurtin2010}
\bibinfo{author}{Gurtin, M.E.}, \bibinfo{author}{Fried, E.},
  \bibinfo{author}{Anand, L.}, \bibinfo{year}{2010}.
\newblock \bibinfo{title}{{The Mechanics and Thermodynamics of continua}}.
\newblock \bibinfo{publisher}{Cambridge University Press},
  \bibinfo{address}{Cambridge, UK}.
\bibitem[{Harris et~al.(2018)Harris, Lawrence, Medlin, Guetard, Burns and
  Somerday}]{Harris2018}
\bibinfo{author}{Harris, Z.D.}, \bibinfo{author}{Lawrence, S.K.},
  \bibinfo{author}{Medlin, D.L.}, \bibinfo{author}{Guetard, G.},
  \bibinfo{author}{Burns, J.T.}, \bibinfo{author}{Somerday, B.P.},
  \bibinfo{year}{2018}.
\newblock \bibinfo{title}{{Elucidating the contribution of mobile
  hydrogen-deformation interactions to hydrogen-induced intergranular cracking
  in polycrystalline nickel}}.
\newblock \bibinfo{journal}{Acta Materialia} \bibinfo{volume}{158},
  \bibinfo{pages}{180--192}.
\bibitem[{Hirshikesh et~al.(2019)Hirshikesh, Natarajan, Annabattula and
  Mart{\'{i}}nez-Pa{\~{n}}eda}]{CPB2019}
\bibinfo{author}{Hirshikesh}, \bibinfo{author}{Natarajan, S.},
  \bibinfo{author}{Annabattula, R.K.},
  \bibinfo{author}{Mart{\'{i}}nez-Pa{\~{n}}eda, E.}, \bibinfo{year}{2019}.
\newblock \bibinfo{title}{{Phase field modelling of crack propagation in
  functionally graded materials}}.
\newblock \bibinfo{journal}{Composites Part B: Engineering}
  \bibinfo{volume}{169}, \bibinfo{pages}{239--248}.
\bibitem[{Hirth(1980)}]{Hirth1980}
\bibinfo{author}{Hirth, J.P.}, \bibinfo{year}{1980}.
\newblock \bibinfo{title}{{Effects of hydrogen on the properties of iron and
  steel}}.
\newblock \bibinfo{journal}{Metallurgical Transactions A} \bibinfo{volume}{11},
  \bibinfo{pages}{861--890}.
\bibitem[{Huang et~al.(2004)Huang, Qu, Hwang, Li, Gao, Huang, Qu, Hwang, Li and
  Gao}]{Huang2004a}
\bibinfo{author}{Huang, Y.}, \bibinfo{author}{Qu, S.}, \bibinfo{author}{Hwang,
  K.C.}, \bibinfo{author}{Li, M.}, \bibinfo{author}{Gao, H.},
  \bibinfo{author}{Huang, Y.}, \bibinfo{author}{Qu, S.},
  \bibinfo{author}{Hwang, K.C.}, \bibinfo{author}{Li, M.},
  \bibinfo{author}{Gao, H.}, \bibinfo{year}{2004}.
\newblock \bibinfo{title}{{A conventional theory of mechanism-based strain
  gradient plasticity}}.
\newblock \bibinfo{journal}{International Journal of Plasticity}
  \bibinfo{volume}{20}, \bibinfo{pages}{753--782}.
\bibitem[{Hutchinson(1983)}]{Hutchinson1983}
\bibinfo{author}{Hutchinson, J.W.}, \bibinfo{year}{1983}.
\newblock \bibinfo{title}{{Fundamentals of the phenomenological theory of
  nonlinear fracture mechanics}}.
\newblock \bibinfo{journal}{Journal of Applied Mechanics, Transactions ASME}
  \bibinfo{volume}{50}, \bibinfo{pages}{1042--1051}.
\bibitem[{Jiang and Carter(2004)}]{Jiang2004a}
\bibinfo{author}{Jiang, D.E.}, \bibinfo{author}{Carter, E.A.},
  \bibinfo{year}{2004}.
\newblock \bibinfo{title}{{First principles assessment of ideal fracture
  energies of materials with mobile impurities: Implications for hydrogen
  embrittlement of metals}}.
\newblock \bibinfo{journal}{Acta Materialia} \bibinfo{volume}{52},
  \bibinfo{pages}{4801--4807}.
\bibitem[{Kehler and Scully(2008)}]{Kehler2008}
\bibinfo{author}{Kehler, B.A.}, \bibinfo{author}{Scully, J.R.},
  \bibinfo{year}{2008}.
\newblock \bibinfo{title}{{Predicting the effect of applied potential on crack
  tip hydrogen concentration in low-alloy martensitic steels}}.
\newblock \bibinfo{journal}{Corrosion} \bibinfo{volume}{64},
  \bibinfo{pages}{465--477}.
\bibitem[{Kiuchi and McLellan(1983)}]{Kiuchi1983}
\bibinfo{author}{Kiuchi, K.}, \bibinfo{author}{McLellan, R.B.},
  \bibinfo{year}{1983}.
\newblock \bibinfo{title}{{The solubility and diffusivity of hydrogen in
  well-annealed and deformed iron}}.
\newblock \bibinfo{journal}{Acta Metallurgica} \bibinfo{volume}{31},
  \bibinfo{pages}{961--984}.
\bibitem[{Kok et~al.(2002)Kok, Beaudoin and Tortorelli}]{Kok2002}
\bibinfo{author}{Kok, S.}, \bibinfo{author}{Beaudoin, A.J.},
  \bibinfo{author}{Tortorelli, D.A.}, \bibinfo{year}{2002}.
\newblock \bibinfo{title}{{A polycrystal plasticity model based on the
  mechanical threshold}}.
\newblock \bibinfo{journal}{International Journal of Plasticity}
  \bibinfo{volume}{18}, \bibinfo{pages}{715--741}.
\bibitem[{Komaragiri et~al.(2008)Komaragiri, Agnew, Gangloff and
  Begley}]{Komaragiri2008}
\bibinfo{author}{Komaragiri, U.}, \bibinfo{author}{Agnew, S.R.},
  \bibinfo{author}{Gangloff, R.P.}, \bibinfo{author}{Begley, M.R.},
  \bibinfo{year}{2008}.
\newblock \bibinfo{title}{{The role of macroscopic hardening and individual
  length-scales on crack tip stress elevation from phenomenological strain
  gradient plasticity}}.
\newblock \bibinfo{journal}{Journal of the Mechanics and Physics of Solids}
  \bibinfo{volume}{56}, \bibinfo{pages}{3527--3540}.
\bibitem[{Kristensen and Mart{\'{i}}nez-Pa{\~{n}}eda(2020)}]{TAFM2020}
\bibinfo{author}{Kristensen, P.K.},
  \bibinfo{author}{Mart{\'{i}}nez-Pa{\~{n}}eda, E.}, \bibinfo{year}{2020}.
\newblock \bibinfo{title}{{Phase field fracture modelling using quasi-Newton
  methods and a new adaptive step scheme}}.
\newblock \bibinfo{journal}{Theoretical and Applied Fracture Mechanics}
  \bibinfo{volume}{107}, \bibinfo{pages}{102446}.
\bibitem[{Kristensen et~al.(2020a)Kristensen, Niordson and
  Mart{\'{i}}nez-Pa{\~{n}}eda}]{JMPS2020}
\bibinfo{author}{Kristensen, P.K.}, \bibinfo{author}{Niordson, C.F.},
  \bibinfo{author}{Mart{\'{i}}nez-Pa{\~{n}}eda, E.}, \bibinfo{year}{2020}a.
\newblock \bibinfo{title}{{A phase field model for elastic-gradient-plastic
  solids undergoing hydrogen embrittlement}}.
\newblock \bibinfo{journal}{Journal of the Mechanics and Physics of Solids}
  \bibinfo{volume}{143}, \bibinfo{pages}{104093}.
\bibitem[{Kristensen et~al.(2020b)Kristensen, Niordson and
  Mart{\'{i}}nez-Pa{\~{n}}eda}]{TAFM2020c}
\bibinfo{author}{Kristensen, P.K.}, \bibinfo{author}{Niordson, C.F.},
  \bibinfo{author}{Mart{\'{i}}nez-Pa{\~{n}}eda, E.}, \bibinfo{year}{2020}b.
\newblock \bibinfo{title}{{Applications of phase field fracture in modelling
  hydrogen assisted failures}}.
\newblock \bibinfo{journal}{Theoretical and Applied Fracture Mechanics}
  \bibinfo{volume}{110}, \bibinfo{pages}{102837}.
\bibitem[{Kumar and Mahajan(2020)}]{Kumar2020}
\bibinfo{author}{Kumar, R.}, \bibinfo{author}{Mahajan, D.K.},
  \bibinfo{year}{2020}.
\newblock \bibinfo{title}{{Hydrogen distribution in metallic polycrystals with
  deformation}}.
\newblock \bibinfo{journal}{Journal of the Mechanics and Physics of Solids}
  \bibinfo{volume}{135}, \bibinfo{pages}{103776}.
\bibitem[{Li et~al.(2004)Li, Gangloff and Scully}]{Li2004}
\bibinfo{author}{Li, D.}, \bibinfo{author}{Gangloff, R.P.},
  \bibinfo{author}{Scully, J.R.}, \bibinfo{year}{2004}.
\newblock \bibinfo{title}{{Hydrogen Trap States in Ultrahigh-Strength AERMET
  100 Steel}}.
\newblock \bibinfo{journal}{Metallurgical and Materials Transactions A:
  Physical Metallurgy and Materials Science} \bibinfo{volume}{35 A},
  \bibinfo{pages}{849--864}.
\bibitem[{Liu et~al.(2005)Liu, Huang, Li, Hwang and Liu}]{Liu2005}
\bibinfo{author}{Liu, B.}, \bibinfo{author}{Huang, Y.}, \bibinfo{author}{Li,
  M.}, \bibinfo{author}{Hwang, K.C.}, \bibinfo{author}{Liu, C.},
  \bibinfo{year}{2005}.
\newblock \bibinfo{title}{{A study of the void size effect based on the Taylor
  dislocation model}}.
\newblock \bibinfo{journal}{International Journal of Plasticity}
  \bibinfo{volume}{21}, \bibinfo{pages}{2107--2122}.
\bibitem[{Lufrano et~al.(1998)Lufrano, Sofronis and Birnbaum}]{Lufrano1998a}
\bibinfo{author}{Lufrano, J.}, \bibinfo{author}{Sofronis, P.},
  \bibinfo{author}{Birnbaum, H.K.}, \bibinfo{year}{1998}.
\newblock \bibinfo{title}{{Elastoplastically accommodated hydride formation and
  embrittlement}}.
\newblock \bibinfo{journal}{Journal of the Mechanics and Physics of Solids}
  \bibinfo{volume}{46}, \bibinfo{pages}{1497--1520}.
\bibitem[{Lynch(2019)}]{Lynch2019}
\bibinfo{author}{Lynch, S.}, \bibinfo{year}{2019}.
\newblock \bibinfo{title}{{Discussion of some recent literature on
  hydrogen-embrittlement mechanisms: Addressing common misunderstandings}}.
\newblock \bibinfo{journal}{Corrosion Reviews} \bibinfo{volume}{37},
  \bibinfo{pages}{377--395}.
\bibitem[{Mart{\'{i}}nez-Pa{\~{n}}eda and Beteg{\'{o}}n(2015)}]{IJSS2015}
\bibinfo{author}{Mart{\'{i}}nez-Pa{\~{n}}eda, E.},
  \bibinfo{author}{Beteg{\'{o}}n, C.}, \bibinfo{year}{2015}.
\newblock \bibinfo{title}{{Modeling damage and fracture within strain-gradient
  plasticity}}.
\newblock \bibinfo{journal}{International Journal of Solids and Structures}
  \bibinfo{volume}{59}, \bibinfo{pages}{208--215}.
\bibitem[{Mart{\'{i}}nez-Pa{\~{n}}eda et~al.(2017)Mart{\'{i}}nez-Pa{\~{n}}eda,
  del Busto and Beteg{\'{o}}n}]{TAFM2017}
\bibinfo{author}{Mart{\'{i}}nez-Pa{\~{n}}eda, E.}, \bibinfo{author}{del Busto,
  S.}, \bibinfo{author}{Beteg{\'{o}}n, C.}, \bibinfo{year}{2017}.
\newblock \bibinfo{title}{{Non-local plasticity effects on notch fracture
  mechanics}}.
\newblock \bibinfo{journal}{Theoretical and Applied Fracture Mechanics}
  \bibinfo{volume}{92}, \bibinfo{pages}{276--287}.
\bibitem[{Mart{\'{i}}nez-Pa{\~{n}}eda et~al.(2019)Mart{\'{i}}nez-Pa{\~{n}}eda,
  Deshpande, Niordson and Fleck}]{JMPS2019}
\bibinfo{author}{Mart{\'{i}}nez-Pa{\~{n}}eda, E.}, \bibinfo{author}{Deshpande,
  V.S.}, \bibinfo{author}{Niordson, C.F.}, \bibinfo{author}{Fleck, N.A.},
  \bibinfo{year}{2019}.
\newblock \bibinfo{title}{{The role of plastic strain gradients in the crack
  growth resistance of metals}}.
\newblock \bibinfo{journal}{Journal of the Mechanics and Physics of Solids}
  \bibinfo{volume}{126}, \bibinfo{pages}{136--150}.
\bibitem[{Mart{\'{i}}nez-Pa{\~{n}}eda et~al.(2020a)Mart{\'{i}}nez-Pa{\~{n}}eda,
  D{\'{i}}az, Wright and Turnbull}]{CS2020b}
\bibinfo{author}{Mart{\'{i}}nez-Pa{\~{n}}eda, E.}, \bibinfo{author}{D{\'{i}}az,
  A.}, \bibinfo{author}{Wright, L.}, \bibinfo{author}{Turnbull, A.},
  \bibinfo{year}{2020}a.
\newblock \bibinfo{title}{{Generalised boundary conditions for hydrogen
  transport at crack tips}}.
\newblock \bibinfo{journal}{Corrosion Science} \bibinfo{volume}{173},
  \bibinfo{pages}{108698}.
\bibitem[{Mart{\'{i}}nez-Pa{\~{n}}eda and Fleck(2019)}]{EJMAS2019}
\bibinfo{author}{Mart{\'{i}}nez-Pa{\~{n}}eda, E.}, \bibinfo{author}{Fleck,
  N.A.}, \bibinfo{year}{2019}.
\newblock \bibinfo{title}{{Mode I crack tip fields: Strain gradient plasticity
  theory versus J2 flow theory}}.
\newblock \bibinfo{journal}{European Journal of Mechanics - A/Solids}
  \bibinfo{volume}{75}, \bibinfo{pages}{381--388}.
\bibitem[{Mart{\'{i}}nez-Pa{\~{n}}eda et~al.(2018)Mart{\'{i}}nez-Pa{\~{n}}eda,
  Golahmar and Niordson}]{CMAME2018}
\bibinfo{author}{Mart{\'{i}}nez-Pa{\~{n}}eda, E.}, \bibinfo{author}{Golahmar,
  A.}, \bibinfo{author}{Niordson, C.F.}, \bibinfo{year}{2018}.
\newblock \bibinfo{title}{{A phase field formulation for hydrogen assisted
  cracking}}.
\newblock \bibinfo{journal}{Computer Methods in Applied Mechanics and
  Engineering} \bibinfo{volume}{342}, \bibinfo{pages}{742--761}.
\bibitem[{Mart{\'{i}}nez-Pa{\~{n}}eda et~al.(2020b)Mart{\'{i}}nez-Pa{\~{n}}eda,
  Harris, Fuentes-Alonso, Scully and Burns}]{CS2020}
\bibinfo{author}{Mart{\'{i}}nez-Pa{\~{n}}eda, E.}, \bibinfo{author}{Harris,
  Z.D.}, \bibinfo{author}{Fuentes-Alonso, S.}, \bibinfo{author}{Scully, J.R.},
  \bibinfo{author}{Burns, J.T.}, \bibinfo{year}{2020}b.
\newblock \bibinfo{title}{{On the suitability of slow strain rate tensile
  testing for assessing hydrogen embrittlement susceptibility}}.
\newblock \bibinfo{journal}{Corrosion Science} \bibinfo{volume}{163},
  \bibinfo{pages}{108291}.
\bibitem[{Mart{\'{i}}nez-Pa{\~{n}}eda and Niordson(2016)}]{IJP2016}
\bibinfo{author}{Mart{\'{i}}nez-Pa{\~{n}}eda, E.}, \bibinfo{author}{Niordson,
  C.F.}, \bibinfo{year}{2016}.
\newblock \bibinfo{title}{{On fracture in finite strain gradient plasticity}}.
\newblock \bibinfo{journal}{International Journal of Plasticity}
  \bibinfo{volume}{80}, \bibinfo{pages}{154--167}.
\bibitem[{Mart{\'{i}}nez-Pa{\~{n}}eda et~al.(2016)Mart{\'{i}}nez-Pa{\~{n}}eda,
  Niordson and Gangloff}]{AM2016}
\bibinfo{author}{Mart{\'{i}}nez-Pa{\~{n}}eda, E.}, \bibinfo{author}{Niordson,
  C.F.}, \bibinfo{author}{Gangloff, R.P.}, \bibinfo{year}{2016}.
\newblock \bibinfo{title}{{Strain gradient plasticity-based modeling of
  hydrogen environment assisted cracking}}.
\newblock \bibinfo{journal}{Acta Materialia} \bibinfo{volume}{117},
  \bibinfo{pages}{321--332}.
\bibitem[{McAuliffe and Waisman(2015)}]{McAuliffe2015}
\bibinfo{author}{McAuliffe, C.}, \bibinfo{author}{Waisman, H.},
  \bibinfo{year}{2015}.
\newblock \bibinfo{title}{{A unified model for metal failure capturing shear
  banding and fracture}}.
\newblock \bibinfo{journal}{International Journal of Plasticity}
  \bibinfo{volume}{65}, \bibinfo{pages}{131--151}.
\bibitem[{Miehe et~al.(2016)Miehe, Aldakheel and Raina}]{Miehe2016b}
\bibinfo{author}{Miehe, C.}, \bibinfo{author}{Aldakheel, F.},
  \bibinfo{author}{Raina, A.}, \bibinfo{year}{2016}.
\newblock \bibinfo{title}{{Phase field modeling of ductile fracture at finite
  strains: A variational gradient-extended plasticity-damage theory}}.
\newblock \bibinfo{journal}{International Journal of Plasticity}
  \bibinfo{volume}{84}, \bibinfo{pages}{1--32}.
\bibitem[{Miehe et~al.(2010)Miehe, Hofacker and Welschinger}]{Miehe2010a}
\bibinfo{author}{Miehe, C.}, \bibinfo{author}{Hofacker, M.},
  \bibinfo{author}{Welschinger, F.}, \bibinfo{year}{2010}.
\newblock \bibinfo{title}{{A phase field model for rate-independent crack
  propagation: Robust algorithmic implementation based on operator splits}}.
\newblock \bibinfo{journal}{Computer Methods in Applied Mechanics and
  Engineering} \bibinfo{volume}{199}, \bibinfo{pages}{2765--2778}.
\bibitem[{Momotani et~al.(2017)Momotani, Shibata, Terada and
  Tsuji}]{Momotani2017}
\bibinfo{author}{Momotani, Y.}, \bibinfo{author}{Shibata, A.},
  \bibinfo{author}{Terada, D.}, \bibinfo{author}{Tsuji, N.},
  \bibinfo{year}{2017}.
\newblock \bibinfo{title}{{Effect of strain rate on hydrogen embrittlement in
  low-carbon martensitic steel}}.
\newblock \bibinfo{journal}{International Journal of Hydrogen Energy}
  \bibinfo{volume}{42}, \bibinfo{pages}{3371--3379}.
\bibitem[{Mu et~al.(2014)Mu, Hutchinson and Meng}]{Mu2014}
\bibinfo{author}{Mu, Y.}, \bibinfo{author}{Hutchinson, J.W.},
  \bibinfo{author}{Meng, W.J.}, \bibinfo{year}{2014}.
\newblock \bibinfo{title}{{Micro-pillar measurements of plasticity in confined
  Cu thin films}}.
\newblock \bibinfo{journal}{Extreme Mechanics Letters} \bibinfo{volume}{1},
  \bibinfo{pages}{62--69}.
\bibitem[{Nix and Gao(1998)}]{Nix1998}
\bibinfo{author}{Nix, W.D.}, \bibinfo{author}{Gao, H.J.}, \bibinfo{year}{1998}.
\newblock \bibinfo{title}{{Indentation size effects in crystalline materials: A
  law for strain gradient plasticity}}.
\newblock \bibinfo{journal}{Journal of the Mechanics and Physics of Solids}
  \bibinfo{volume}{46}, \bibinfo{pages}{411--425}.
\bibitem[{Oriani and Josephic(1974)}]{Oriani1974}
\bibinfo{author}{Oriani, R.A.}, \bibinfo{author}{Josephic, P.H.},
  \bibinfo{year}{1974}.
\newblock \bibinfo{title}{{Equilibrium Aspects of Hydrogen Induced Cracking of
  Steels}}.
\newblock \bibinfo{journal}{Acta Metallurgica} \bibinfo{volume}{22},
  \bibinfo{pages}{1065--1074}.
\bibitem[{Orowan(1948)}]{Orowan1948}
\bibinfo{author}{Orowan, E.}, \bibinfo{year}{1948}.
\newblock \bibinfo{title}{{Fracture and Strength of Solids}}.
\newblock \bibinfo{journal}{Reports on Progress in Physics}
  \bibinfo{volume}{XII}, \bibinfo{pages}{185}.
\bibitem[{Papazafeiropoulos et~al.(2017)Papazafeiropoulos, Mu{\~{n}}iz-Calvente
  and Mart{\'{i}}nez-Pa{\~{n}}eda}]{AES2017}
\bibinfo{author}{Papazafeiropoulos, G.}, \bibinfo{author}{Mu{\~{n}}iz-Calvente,
  M.}, \bibinfo{author}{Mart{\'{i}}nez-Pa{\~{n}}eda, E.}, \bibinfo{year}{2017}.
\newblock \bibinfo{title}{{Abaqus2Matlab: A suitable tool for finite element
  post-processing}}.
\newblock \bibinfo{journal}{Advances in Engineering Software}
  \bibinfo{volume}{105}, \bibinfo{pages}{9--16}.
\bibitem[{Paxton et~al.(2017)Paxton, Sutton and Finnis}]{Paxton2017}
\bibinfo{author}{Paxton, T.}, \bibinfo{author}{Sutton, A.P.},
  \bibinfo{author}{Finnis, M.W.}, \bibinfo{year}{2017}.
\newblock \bibinfo{title}{{The challenges of hydrogen and metals}}.
\newblock \bibinfo{journal}{Philosophical Transactions of the Royal Society A:
  Mathematical, Physical and Engineering Sciences} \bibinfo{volume}{375}.
\bibitem[{Pouillier et~al.(2012)Pouillier, Gourgues, Tanguy and
  Busso}]{Pouillier2012}
\bibinfo{author}{Pouillier, E.}, \bibinfo{author}{Gourgues, A.F.},
  \bibinfo{author}{Tanguy, D.}, \bibinfo{author}{Busso, E.P.},
  \bibinfo{year}{2012}.
\newblock \bibinfo{title}{{A study of intergranular fracture in an aluminium
  alloy due to hydrogen embrittlement}}.
\newblock \bibinfo{journal}{International Journal of Plasticity}
  \bibinfo{volume}{34}, \bibinfo{pages}{139--153}.
\bibitem[{Provatas and Elder(2011)}]{Provatas2011}
\bibinfo{author}{Provatas, N.}, \bibinfo{author}{Elder, K.},
  \bibinfo{year}{2011}.
\newblock \bibinfo{title}{{Phase-Field Methods in Materials Science and
  Engineering}}.
\newblock \bibinfo{publisher}{John Wiley {\&} Sons},
  \bibinfo{address}{Weinheim, Germany}.
\bibitem[{Qu et~al.(2004)Qu, Huang, Jiang, Liu, Wu and Hwang}]{Qu2004}
\bibinfo{author}{Qu, S.}, \bibinfo{author}{Huang, Y.}, \bibinfo{author}{Jiang,
  H.}, \bibinfo{author}{Liu, C.}, \bibinfo{author}{Wu, P.D.},
  \bibinfo{author}{Hwang, K.C.}, \bibinfo{year}{2004}.
\newblock \bibinfo{title}{{Fracture analysis in the conventional theory of
  mechanism-based strain gradient (CMSG) plasticity}}.
\newblock \bibinfo{journal}{International Journal of Fracture}
  \bibinfo{volume}{129}, \bibinfo{pages}{199--220}.
\bibitem[{Quintanas-Corominas et~al.(2020)Quintanas-Corominas, Turon, Reinoso,
  Casoni, Paggi and Mayugo}]{Quintanas-Corominas2020a}
\bibinfo{author}{Quintanas-Corominas, A.}, \bibinfo{author}{Turon, A.},
  \bibinfo{author}{Reinoso, J.}, \bibinfo{author}{Casoni, E.},
  \bibinfo{author}{Paggi, M.}, \bibinfo{author}{Mayugo, J.A.},
  \bibinfo{year}{2020}.
\newblock \bibinfo{title}{{A phase field approach enhanced with a cohesive zone
  model for modeling delamination induced by matrix cracking}}.
\newblock \bibinfo{journal}{Computer Methods in Applied Mechanics and
  Engineering} \bibinfo{volume}{358}, \bibinfo{pages}{112618}.
\bibitem[{Serebrinsky et~al.(2004)Serebrinsky, Carter and
  Ortiz}]{Serebrinsky2004}
\bibinfo{author}{Serebrinsky, S.}, \bibinfo{author}{Carter, E.A.},
  \bibinfo{author}{Ortiz, M.}, \bibinfo{year}{2004}.
\newblock \bibinfo{title}{{A quantum-mechanically informed continuum model of
  hydrogen embrittlement}}.
\newblock \bibinfo{journal}{Journal of the Mechanics and Physics of Solids}
  \bibinfo{volume}{52}, \bibinfo{pages}{2403--2430}.
\bibitem[{Shi et~al.(2001)Shi, Huang, Jiang, Hwang and Li}]{Shi2001}
\bibinfo{author}{Shi, M.}, \bibinfo{author}{Huang, Y.}, \bibinfo{author}{Jiang,
  H.}, \bibinfo{author}{Hwang, K.C.}, \bibinfo{author}{Li, M.},
  \bibinfo{year}{2001}.
\newblock \bibinfo{title}{{The boundary-layer effect on the crack tip field in
  mechanism-based strain gradient plasticity}}.
\newblock \bibinfo{journal}{International Journal of Fracture}
  \bibinfo{volume}{112}, \bibinfo{pages}{23--41}.
\bibitem[{Shi et~al.(2004)Shi, Huang and Gao}]{Shi2004}
\bibinfo{author}{Shi, M.X.}, \bibinfo{author}{Huang, Y.}, \bibinfo{author}{Gao,
  H.}, \bibinfo{year}{2004}.
\newblock \bibinfo{title}{{The J-integral and geometrically necessary
  dislocations in nonuniform plastic deformation}}.
\newblock \bibinfo{journal}{International Journal of Plasticity}
  \bibinfo{volume}{20}, \bibinfo{pages}{1739--1762}.
\bibitem[{Shishvan et~al.(2020)Shishvan, Cs{\'{a}}nyi and
  Deshpande}]{Shishvan2020}
\bibinfo{author}{Shishvan, S.S.}, \bibinfo{author}{Cs{\'{a}}nyi, G.},
  \bibinfo{author}{Deshpande, V.S.}, \bibinfo{year}{2020}.
\newblock \bibinfo{title}{{Hydrogen induced fast-fracture}}.
\newblock \bibinfo{journal}{Journal of the Mechanics and Physics of Solids}
  \bibinfo{volume}{134}, \bibinfo{pages}{103740}.
\bibitem[{Simoes and Mart{\'{i}}nez-Pa{\~{n}}eda(2021)}]{CMAME2021}
\bibinfo{author}{Simoes, M.}, \bibinfo{author}{Mart{\'{i}}nez-Pa{\~{n}}eda,
  E.}, \bibinfo{year}{2021}.
\newblock \bibinfo{title}{{Phase field modelling of fracture and fatigue in
  Shape Memory Alloys}}.
\newblock \bibinfo{journal}{Computer Methods in Applied Mechanics and
  Engineering} \bibinfo{volume}{373}, \bibinfo{pages}{113504}.
\bibitem[{Sofronis et~al.(2001)Sofronis, Liang and Aravas}]{Sofronis2001}
\bibinfo{author}{Sofronis, P.}, \bibinfo{author}{Liang, Y.},
  \bibinfo{author}{Aravas, N.}, \bibinfo{year}{2001}.
\newblock \bibinfo{title}{{Hydrogen induced shear localization of the plastic
  flow in metals and alloys}}.
\newblock \bibinfo{journal}{Eur. J. Mech. A/Solids} \bibinfo{volume}{20},
  \bibinfo{pages}{857--872}.
\bibitem[{Sofronis and McMeeking(1989)}]{Sofronis1989}
\bibinfo{author}{Sofronis, P.}, \bibinfo{author}{McMeeking, R.M.},
  \bibinfo{year}{1989}.
\newblock \bibinfo{title}{{Numerical analysis of hydrogen transport near a
  blunting crack tip}}.
\newblock \bibinfo{journal}{Journal of the Mechanics and Physics of Solids}
  \bibinfo{volume}{37}, \bibinfo{pages}{317--350}.
\bibitem[{Tann{\'{e}} et~al.(2018)Tann{\'{e}}, Li, Bourdin, Marigo and
  Maurini}]{Tanne2018}
\bibinfo{author}{Tann{\'{e}}, E.}, \bibinfo{author}{Li, T.},
  \bibinfo{author}{Bourdin, B.}, \bibinfo{author}{Marigo, J.J.},
  \bibinfo{author}{Maurini, C.}, \bibinfo{year}{2018}.
\newblock \bibinfo{title}{{Crack nucleation in variational phase-field models
  of brittle fracture}}.
\newblock \bibinfo{journal}{Journal of the Mechanics and Physics of Solids}
  \bibinfo{volume}{110}, \bibinfo{pages}{80--99}.
\bibitem[{Taylor(1938)}]{Taylor1938}
\bibinfo{author}{Taylor, G.I.}, \bibinfo{year}{1938}.
\newblock \bibinfo{title}{{Plastic strain in metals}}.
\newblock \bibinfo{journal}{Journal of the Institute of Metals}
  \bibinfo{volume}{62}, \bibinfo{pages}{307--324}.
\bibitem[{Turnbull(2015)}]{Turnbull2015}
\bibinfo{author}{Turnbull, A.}, \bibinfo{year}{2015}.
\newblock \bibinfo{title}{{Perspectives on hydrogen uptake, diffusion and
  trapping}}.
\newblock \bibinfo{journal}{International Journal of Hydrogen Energy}
  \bibinfo{volume}{40}, \bibinfo{pages}{16961--16970}.
\bibitem[{Tvergaard and Hutchinson(1992)}]{Tvergaard1992}
\bibinfo{author}{Tvergaard, V.}, \bibinfo{author}{Hutchinson, J.W.},
  \bibinfo{year}{1992}.
\newblock \bibinfo{title}{{The relation between crack growth resistance and
  fracture process parameters in elastic-plastic solids}}.
\newblock \bibinfo{journal}{Journal of the Mechanics and Physics of Solids}
  \bibinfo{volume}{40}, \bibinfo{pages}{1377--1397}.
\bibitem[{Tvergaard and Niordson(2004)}]{Tvergaard2004a}
\bibinfo{author}{Tvergaard, V.}, \bibinfo{author}{Niordson, C.F.},
  \bibinfo{year}{2004}.
\newblock \bibinfo{title}{{Nonlocal plasticity effects on interaction of
  different size voids}}.
\newblock \bibinfo{journal}{International Journal of Plasticity}
  \bibinfo{volume}{20}, \bibinfo{pages}{107--120}.
\bibitem[{{Van der Ven} and Ceder(2003)}]{VanderVen2003}
\bibinfo{author}{{Van der Ven}, A.}, \bibinfo{author}{Ceder, G.},
  \bibinfo{year}{2003}.
\newblock \bibinfo{title}{{Impurity-induced van der Waals transition during
  decohesion}}.
\newblock \bibinfo{journal}{Physical Review B - Condensed Matter and Materials
  Physics} \bibinfo{volume}{67}, \bibinfo{pages}{1--4}.
\bibitem[{{Van Leeuwen}(1974)}]{VanLeeuwen1974}
\bibinfo{author}{{Van Leeuwen}, H.P.}, \bibinfo{year}{1974}.
\newblock \bibinfo{title}{{The kinetics of hydrogen embrittlement: A
  quantitative diffusion model}}.
\newblock \bibinfo{journal}{Engineering Fracture Mechanics}
  \bibinfo{volume}{6}, \bibinfo{pages}{141--161}.
\bibitem[{Voyiadjis and Song(2019)}]{Voyiadjis2019}
\bibinfo{author}{Voyiadjis, G.Z.}, \bibinfo{author}{Song, Y.},
  \bibinfo{year}{2019}.
\newblock \bibinfo{title}{{Strain gradient continuum plasticity theories:
  Theoretical, numerical and experimental investigations}}.
\newblock \bibinfo{journal}{International Journal of Plasticity}
  \bibinfo{volume}{121}, \bibinfo{pages}{21--75}.
\bibitem[{Wang et~al.(2005)Wang, Akiyama and Tsuzaki}]{Wang2005}
\bibinfo{author}{Wang, M.}, \bibinfo{author}{Akiyama, E.},
  \bibinfo{author}{Tsuzaki, K.}, \bibinfo{year}{2005}.
\newblock \bibinfo{title}{{Effect of hydrogen and stress concentration on the
  notch tensile strength of AISI 4135 steel}}.
\newblock \bibinfo{journal}{Materials Science and Engineering A}
  \bibinfo{volume}{398}, \bibinfo{pages}{37--46}.
\bibitem[{Wei and Hutchinson(1997)}]{Wei1997}
\bibinfo{author}{Wei, Y.}, \bibinfo{author}{Hutchinson, J.W.},
  \bibinfo{year}{1997}.
\newblock \bibinfo{title}{{Steady-state crack growth and work of fracture for
  solids characterized by strain gradient plasticity}}.
\newblock \bibinfo{journal}{Journal of the Mechanics and Physics of Solids}
  \bibinfo{volume}{45}, \bibinfo{pages}{1253--1273}.
\bibitem[{Wei and Xu(2005)}]{Wei2005}
\bibinfo{author}{Wei, Y.}, \bibinfo{author}{Xu, G.}, \bibinfo{year}{2005}.
\newblock \bibinfo{title}{{A multiscale model for the ductile fracture of
  crystalline materials}}.
\newblock \bibinfo{journal}{International Journal of Plasticity}
  \bibinfo{volume}{21}, \bibinfo{pages}{2123--2149}.
\bibitem[{Williams(1957)}]{Williams1957}
\bibinfo{author}{Williams, M.L.}, \bibinfo{year}{1957}.
\newblock \bibinfo{title}{{On the stress distribution at the base of a
  stationary crack}}.
\newblock \bibinfo{journal}{Journal of Applied Mechanics} \bibinfo{volume}{24},
  \bibinfo{pages}{109--114}.
\bibitem[{Wu et~al.(2020a)Wu, Mandal and Nguyen}]{Wu2020b}
\bibinfo{author}{Wu, J.Y.}, \bibinfo{author}{Mandal, T.K.},
  \bibinfo{author}{Nguyen, V.P.}, \bibinfo{year}{2020}a.
\newblock \bibinfo{title}{{A phase-field regularized cohesive zone model for
  hydrogen assisted cracking}}.
\newblock \bibinfo{journal}{Computer Methods in Applied Mechanics and
  Engineering} \bibinfo{volume}{358}, \bibinfo{pages}{112614}.
\bibitem[{Wu et~al.(2020b)Wu, Nguyen, Nguyen, Sutula, Sinaie and
  Bordas}]{Wu2020}
\bibinfo{author}{Wu, J.Y.}, \bibinfo{author}{Nguyen, V.P.},
  \bibinfo{author}{Nguyen, C.T.}, \bibinfo{author}{Sutula, D.},
  \bibinfo{author}{Sinaie, S.}, \bibinfo{author}{Bordas, S.},
  \bibinfo{year}{2020}b.
\newblock \bibinfo{title}{{Phase-field modelling of fracture}}.
\newblock \bibinfo{journal}{Advances in Applied Mechanics}
  \bibinfo{volume}{53}, \bibinfo{pages}{1--183}.
\bibitem[{You et~al.(2021)You, Waisman and Zhu}]{You2021}
\bibinfo{author}{You, T.}, \bibinfo{author}{Waisman, H.}, \bibinfo{author}{Zhu,
  Q.Z.}, \bibinfo{year}{2021}.
\newblock \bibinfo{title}{{Brittle-ductile failure transition in geomaterials
  modeled by a modified phase-field method with a varying damage-driving energy
  coefficient}}.
\newblock \bibinfo{journal}{International Journal of Plasticity}
  \bibinfo{volume}{136}, \bibinfo{pages}{102836}.
\bibitem[{Yu et~al.(2016)Yu, Olsen, Alvaro, Olden, He and Zhang}]{Yu2016a}
\bibinfo{author}{Yu, H.}, \bibinfo{author}{Olsen, J.S.},
  \bibinfo{author}{Alvaro, A.}, \bibinfo{author}{Olden, V.},
  \bibinfo{author}{He, J.}, \bibinfo{author}{Zhang, Z.}, \bibinfo{year}{2016}.
\newblock \bibinfo{title}{{A uniform hydrogen degradation law for high strength
  steels}}.
\newblock \bibinfo{journal}{Engineering Fracture Mechanics}
  \bibinfo{volume}{157}, \bibinfo{pages}{56--71}.

\end{thebibliography}

\end{document}